\documentclass[times,sort&compress,3p]{elsarticle}
\usepackage[labelfont=bf]{caption}

\usepackage{amsmath,amsfonts,amssymb,amsthm,booktabs,color,epsfig,graphicx,hyperref,url,enumitem,algorithm,algpseudocode,subcaption,caption}

\theoremstyle{plain}

\newtheorem{proposition}{Proposition}

\theoremstyle{definition}
\newtheorem{definition}{Definition}

\begin{document}

\begin{frontmatter}

\title{The multirank likelihood for semiparametric canonical correlation analysis}

\author{Jordan G. Bryan$^1$ \corref{mycorrespondingauthor}}
\author{Jonathan Niles-Weed$^2$}
\author{Peter D. Hoff$^1$}

\address{$^1$Department of Statistical Science, Duke University, Durham, NC 27708}
\address{$^2$Courant Institute of Mathematical Sciences and the Center for Data Science, NYU, New York, NY 10012}

\cortext[mycorrespondingauthor]{Corresponding author. Email address: \url{jordan.bryan@alumni.duke.edu}}

\begin{abstract}
Many analyses of multivariate data focus on evaluating the dependence between two sets of variables, rather than the dependence among individual variables within each set. Canonical correlation analysis (CCA) is a classical data analysis technique that estimates parameters describing the dependence between such sets. However, inference procedures based on traditional CCA rely on the assumption that all variables are jointly normally distributed. We present a semiparametric approach to CCA in which the multivariate margins of each variable set may be arbitrary, but the dependence between variable sets is described by a parametric model that provides low-dimensional summaries of dependence. While maximum likelihood estimation in the proposed model is intractable, we propose two estimation strategies: one using a pseudolikelihood for the model and one using a Markov chain Monte Carlo (MCMC) algorithm that provides Bayesian estimates and confidence regions for the between-set dependence parameters. The MCMC algorithm is derived from a multirank likelihood function, which uses only part of the information in the observed data in exchange for being free of assumptions about the multivariate margins. We apply the proposed Bayesian inference procedure to Brazilian climate data and monthly stock returns from the materials and communications market sectors.
\end{abstract}


\end{frontmatter}

\section{Introduction\label{sec:1}}

\label{sec:intro}

Scientific studies often involve the collection of multivariate data with complex interdependencies. In some cases, a study will be concerned with the pairwise dependence between individual variables. In other cases, however, the dependence of interest is between non-overlapping sets of variables. For instance, when analyzing financial data, it may be of interest to characterize the association between market sectors without regards to the dependence among stocks within a market sector. Other examples are encountered in biological studies concerning the relationship between diet and the human microbiome \citep{chen_structure-constrained_2013}, psychological studies comparing attachment types to personality disorders \citep{sherry_conducting_2005}, and neuroscience studies comparing brain imaging results to non-imaging measurements (see e.g. \cite{winkler_permutation_2020} for a survey).

One approach to characterizing the association between sets of variables is to conduct hypothesis tests for independence. Classical methods using Wilks' $\Lambda$ statistic \citep{mardia_multivariate_1979}, as well as modern methods using the distance covariance \citep{bakirov_multivariate_2006}, and the Ball covariance \citep{pan_ball_2020} are available for this purpose. While each of these tests relies on some assumptions about the multivariate marginal distributions of the variable sets, recently \cite{shi_distribution-free_2020} and \cite{deb_multivariate_2021} have proposed tests for independence that allow the marginal distributions of the variable sets to be arbitrary. In analogy to tests for independence based on univariate ranks, these tests are based on the multivariate ranks introduced by \cite{chernozhukov_mongekantorovich_2017}, which are defined using the theory of optimal transport.

The multivariate ranks described by these authors have motivated a number of recent works studying the properties of transport-based multivariate counterparts to univariate concepts. For example \cite{hallin_distribution_2021, ghosal_multivariate_2022} characterize notions of multivariate rank and quantile functions using optimal transport criteria, and derive consistency properties of their estimates. In addition to the independence tests mentioned above, \cite{deb_multivariate_2021} also derives a test for equality in distribution, while \cite{huang_multivariate_2023} proposes tests for several forms of multivariate symmetry. In \cite{hallin_rank-based_2023}, the authors extend these ideas to independence testing between successive realizations of random vectors arising from an autoregressive process.

Multivariate ranks based on optimal transport have inspired many recent developments because test statistics based on these ranks yield approximate or exact null distributions that do not assume anything about the multivariate distributions of the variable sets. However, model-based approaches using these ranks are less developed in the literature, and they may have some advantages compared to non-parametric tests. For example, some models have parameters that provide low-dimensional summaries of the association between the variable sets, which may be scientifically interpretable. Canonical correlation analysis (CCA) is a classical data analysis technique that estimates such parameters. First described by \cite{hotelling_relations_1936}, CCA can be motivated from several perspectives, including invariance with respect to transformations under the general linear group \citep{eaton_multivariate_1983}, estimation in a latent factor model \citep{bach_probabilistic_2005}, and statistical whitening \citep{jendoubi_whitening_2019}. Given random vectors $\boldsymbol{y}_1, \boldsymbol{y}_2$ taking values in $\mathbb{R}^{p_1}, \mathbb{R}^{p_2}$, CCA identifies a pair of linear transformations $\boldsymbol{B}_1, \boldsymbol{B}_2$ such that the \textit{canonical variables} $\boldsymbol{x}_1 = \boldsymbol{B}^\top_1 \boldsymbol{y}_1$ and $\boldsymbol{x}_2 = \boldsymbol{B}^\top_2 \boldsymbol{y}_2$  satisfy
\begin{align*} 
   \mathrm{Var}(\boldsymbol{x}_1 ) =  {\bf I}_{p_1}, ~\mathrm{Var}(\boldsymbol{x}_2 ) =  {\bf I}_{p_2}, ~\mathrm{Cov}(\boldsymbol{x}_1, \boldsymbol{x}_2) = \boldsymbol{\Lambda},
\end{align*}
where $\boldsymbol{\Lambda}$ is a diagonal matrix with ordered diagonal entries $1 > \lambda_{1} \geq \dots \geq \lambda_{d} \geq 0$, which are known as the \textit{canonical correlations}. The ordering of the canonical correlations implies, for example, that the correlation of the first element of $\boldsymbol{x}_1$ with that of $\boldsymbol{x}_2$ is the largest among correlations between linear combinations of $\boldsymbol y_{1}$ with those 
of $\boldsymbol y_{2}$. Because they are a function of the original variables, the canonical variables may be useful in exploratory analyses where the goal is to understand which of the original variables is contributing to the association between variable sets. Additionally, the canonical correlations might be helpful in determining the effective dimensionality of the between-set association. However, estimation and uncertainty quantification for the parameters of CCA is challenging without the restrictive assumption that the variables from each set are jointly normally distributed.

In this article, we develop a semiparametric approach to CCA, which preserves the parametric model for between-set dependence, but allows the multivariate margins of each variable set to be arbitrary. Our model extends existing proposals for semiparametric CCA \citep{zoh_pcan_2016, agniel_analysis_2017, yoon_sparse_2020}, which assume that the multivariate marginal distributions of the variable sets can be described by a Gaussian copula. In fact, our model may be seen as a generalization of the Gaussian copula model to vector-valued margins, much like the vector copula introduced by \cite{fan_vector_2023}. Here, though, we present an inference strategy that allows for estimation of and uncertainty quantification for parameters describing the association between variable sets, even when the transformations parameterizing the multivariate margins are unknown. Our inference strategy is based on a multirank likelihood, which uses only part of the information in the observed data in exchange for being free of assumptions about the multivariate margins. In particular, the multirank likelihood uses a notion of ordering in multiple dimensions, which is inspired by the optimal transport criteria of \cite{chernozhukov_mongekantorovich_2017}. While maximum likelihood estimation with the multirank likelihood is intractable, we show that Bayesian estimation of the between-set dependence parameters can be achieved with an MCMC algorithm that simulates from the posterior distribution of the CCA parameters. 

In the first part of Section 2 of this article, we describe a CCA parameterization of the multivariate normal model for variable sets, which separates the parameters describing between-set dependence from those determining the multivariate marginal distributions of the variable sets. We then introduce our model for semiparametric CCA, a Gaussian transformation model whose multivariate margins are parameterized by cyclically monotone functions. In Section 3, we define the multirank likelihood and use it to develop a Bayesian inference strategy for obtaining estimates and confidence regions for the CCA parameters. We then discuss the details of the MCMC algorithm allowing us to simulate from the posterior distribution of the CCA parameters. In Section 4 we illustrate the use of our model for semiparametric CCA on simulated datasets and apply the model to two real datasets: one containing measurements of climate variables in Brazil, and one containing monthly stock returns from the materials and communications market sectors. We conclude with a discussion of possible extensions to this work in Section 5. By default, roman characters referring to mathematical objects in this article are italicized. However, where necessary, we use italicized and un-italicized roman characters to distinguish between random variables and elements of their sample spaces.

\section{Semiparametric CCA}
Let $\boldsymbol Y$ be a random mean-zero $n\times (p_1 + p_2)$ data matrix with independent rows, with the first $p_1$ columns given by the $n\times p_1$ matrix $\boldsymbol Y_1$ and the last 
$p_2$ columns given by the $n\times p_2$ matrix $\boldsymbol Y_2$. The columns of $\boldsymbol Y_1$ and of $\boldsymbol Y_2$ represent two separate variable sets, which may be associated in some manner. One model to evaluate the dependence between the variable sets is the multivariate normal model,
\begin{equation} 
\boldsymbol Y = 
             [ \boldsymbol{Y}_{1}    \ 
             \boldsymbol{Y}_{2} ]
        \sim N_{n\times (p_1 + p_2)}\left( \boldsymbol{0},
        \boldsymbol{\Sigma} \otimes \boldsymbol I_n \right ) \ ,
\end{equation} 
where $\boldsymbol{\Sigma}$ is an unknown positive definite matrix and 
``$\otimes$'' is the Kronecker product. 
Here, the association between 
variable sets $\boldsymbol{Y}_1$ and $\boldsymbol{Y}_2$ may be evaluated by parameterizing $\boldsymbol{\Sigma}$ as  
 \begin{align} 
    \boldsymbol{\Sigma} & =     \left[\begin{array}{cc}
        \boldsymbol{\Sigma}_{1} & \boldsymbol{\Omega}  \\
        \boldsymbol{\Omega}^\top & \boldsymbol{\Sigma}_{2}
    \end{array}\right] 
\end{align} 
where $\boldsymbol{\Sigma}_1$ and $\boldsymbol{\Sigma}_2$ 
are the marginal covariance matrices of $\boldsymbol{Y}_1$ 
and $\boldsymbol{Y}_2$, respectively. 
This parameterization of the multivariate normal model is written so that the covariance matrix is partitioned into $p_1$- and $p_2$-dimensional blocks, and the cross-covariance is dentoted by a $p_1 \times p_2$ matrix parameter $\boldsymbol{\Omega}$. Note that $\boldsymbol{\Omega}$ characterizes the dependence between variable sets, but it is also implicitly a function of $\boldsymbol{\Sigma}_{1}$ and $\boldsymbol{\Sigma}_{2}$ because the joint covariance matrix $\boldsymbol{\Sigma}$ must be positive definite. Our objective is to evaluate the dependence between two sets of variables without respect to the dependence within variable sets, so we work instead with the \textit{CCA parameterization} of the multivariate normal distribution, which we express as
\begin{align}\label{eq:norm_mod} 
 \boldsymbol Z & = [
             \boldsymbol{Z}_{1}   \
             \boldsymbol{Z}_{2} ]
       \sim N_{n\times (p_1 + p_2)}\left( \boldsymbol{0}, \left[\begin{array}{cc}
        \boldsymbol{I}_{p_1} & \boldsymbol{Q}_1 \boldsymbol{\Lambda} \boldsymbol{Q}_2^\top  \\
        \boldsymbol{Q}_2 \boldsymbol{\Lambda} \boldsymbol{Q}_1^\top & \boldsymbol{I}_{p_2}
    \end{array}\right]\otimes \boldsymbol I_n\right) ,   \\
  \boldsymbol{y}_i & = \begin{bmatrix} \boldsymbol{y}_{i,1} \\ 
       \boldsymbol{y}_{i,2} \end{bmatrix}   =  
        \begin{bmatrix} \boldsymbol{\Sigma}_1^{1/2} \boldsymbol{z}_{i,1} \\ 
 \boldsymbol{\Sigma}_2^{1/2} \boldsymbol{z}_{i,2} \end{bmatrix}  
  \nonumber
\end{align}
where $\boldsymbol{y}_{i,j}$ and $\boldsymbol{z}_{i,j}$ are the $i^\text{th}$ 
rows of $\boldsymbol{Y}_j$ and $\boldsymbol{Z}_j$, $j\in \{1,2\}$. 
Here, $\boldsymbol{Q}_1 \in \mathbb{R}^{p_1 \times d}$, $\boldsymbol{Q}_2 \in \mathbb{R}^{p_2 \times d}$ are orthogonal matrices, $\boldsymbol{\Lambda} \in \mathbb{R}^{d \times d}$ is a diagonal matrix with decreasing entries in $[0, 1)$, and $d = \min(p_1, p_2)$. Going forward, we refer to $(\boldsymbol{Q}_1, \boldsymbol{Q}_2, \boldsymbol{\Lambda})$ as the \textit{CCA parameters}, as they neither determine nor depend on $\boldsymbol{\Sigma}_1$ and $\boldsymbol{\Sigma}_2$, the marginal covariances of $\boldsymbol{Y}_1$ 
and $\boldsymbol{Y}_2$. Conversely, the $p_1$-variate marginal distribution 
of $\boldsymbol{Y}_1$ and the $p_2$-variate marginal distribution of 
$\boldsymbol{Y}_2$ are completely determined by 
$\boldsymbol{\Sigma}_1$ and $\boldsymbol{\Sigma}_2$, respectively, 
and do not depend on the CCA parameters. 

The distribution theory of traditional CCA \citep{mardia_multivariate_1979, anderson_asymptotic_1999} is based on the multivariate normal model for $[ \boldsymbol{Y}_{1} \   \boldsymbol{Y}_{2}]$, in which the CCA parameters are the estimands. As the CCA parameterization of the multivariate normal model makes clear, there are two aspects to this model: a multivariate normal model for between-set dependence and a pair of linear transformations parameterizing the multivariate margins of the variable sets. While the normal dependence model is appealing because of the availability of straightforward inferential methods, the assumption that the transformations are linear is restrictive. In particular, this assumption is inappropriate for the analysis of many multivariate datasets, such as those containing data with restricted range, data of mixed type, or other data whose joint distribution is not approximately multivariate normal. 
Our proposal for semiparametric CCA is therefore to expand the class of marginal transformations to a larger set of functions, large enough to accommodate any pair of 
marginal 
distributions on $\mathbb R^{p_1}$ and $\mathbb R^{p_2}$. Specifically, our semiparametric CCA model is
\begin{align}\label{eq:cmcca_mod} 
 \boldsymbol Z & = [
             \boldsymbol{Z}_{1}   \
             \boldsymbol{Z}_{2} ]
       \sim N_{n\times (p_1 + p_2)}\left( \boldsymbol{0}, \left[\begin{array}{cc}
        \boldsymbol{I}_{p_1} & \boldsymbol{Q}_1 \boldsymbol{\Lambda} \boldsymbol{Q}_2^\top  \\
        \boldsymbol{Q}_2 \boldsymbol{\Lambda} \boldsymbol{Q}_1^\top & \boldsymbol{I}_{p_2}
    \end{array}\right]\otimes \boldsymbol I_n\right) ,   \\
  \boldsymbol{y}_i & = \begin{bmatrix} \boldsymbol{y}_{i,1} \\ 
       \boldsymbol{y}_{i,2} \end{bmatrix}   =  
        \begin{bmatrix} G_1(\boldsymbol{z}_{i,1}) \\ 
     G_2(\boldsymbol{z}_{i,2})\end{bmatrix}  
  \nonumber, 
\end{align}
with the model parameters being the CCA parameters 
$(\boldsymbol{Q}_1, \boldsymbol{Q}_2,\boldsymbol{\Lambda})$ and the 
unknown transformations $G_1\in \mathcal G^{p_1}$ and 
$G_2\in \mathcal G^{p_2}$
that determine
the marginal distributions of $\boldsymbol{y}_{i,1}$ 
and $\boldsymbol{y}_{i,2}$, respectively. 

The sets
$\mathcal G^{p_1}$ and $\mathcal G^{p_2}$ of possible values of 
$G_1$ and $G_2$ should be large enough to 
allow the marginal distributions of $\boldsymbol{y}_{i,1}$ and 
$\boldsymbol{y}_{i,2}$ to be arbitrary, but not so large that 
non-identifiability prohibits inference on the CCA parameters. For this reason, we parameterize our model so that 
 $\mathcal{G}^{p_j}$ is the class of \textit{cyclically monotone} functions on $\mathbb{R}^{p_j}$, which is defined with respect to the following criterion.
\begin{definition}[Cyclical monotonicity]\label{def:cm}
    A subset $S$ of $\mathbb{R}^p \times \mathbb{R}^p$ is \textit{cyclically monotone} if and only if
    \begin{equation}\label{eq:cm}
        \sum_{i=1}^n  \mathbf{z}_i^\top \mathbf{y}_i  \geq \sum_{i=1}^n \mathbf{z}_i^\top \mathbf{y}_{\sigma(i)}
    \end{equation}
    for any finite collection of points $\left\{ (\mathbf{z}_1, \mathbf{y}_1), \dots, (\mathbf{z}_n, \mathbf{y}_n) \right\} \subseteq S$ and any permutation $\sigma$.
\end{definition}
Stating that a function $G : \mathbb{R}^p \rightarrow \mathbb{R}^p$ is cyclically monotone means that its graph, the set of all pairs $(\mathbf{z}, G(\mathbf{z}))$, is cyclically monotone. The next two propositions show that  
a model for semiparametric CCA parameterized by cyclically monotone functions achieves the desired balance of flexibility and identifiability.
\begin{proposition}\label{prop:p1}
Let $P$ be a probability distribution on $\mathbb{R}^{p}$ and let $\boldsymbol{z} \sim N_p(\boldsymbol{0}, \boldsymbol{I}_{p})$. Further, let
\begin{align*}
\mathcal{G}^p = \left\{ G~|~ G : \mathbb{R}^{p} \rightarrow \mathbb{R}^{p} \text{ is cyclically monotone} \right\}.
\end{align*}
Then there exists a unique $G\in \mathcal{G}^p$ such that $\boldsymbol{y} = G(\boldsymbol{z}) \sim P$.
\end{proposition}

\begin{proof}[Proof of Proposition 1]
The main theorems of \cite{brenier_polar_1991} and \cite{mccann_existence_1995} assert the existence of a convex function $\psi$, unique almost everywhere up to addition of a constant, such that $\nabla \psi$ pushes forward the standard normal distribution on $\mathbb{R}^{p}$ to $P$. Therefore, there exists an almost-everywhere unique $\nabla \psi$ such that
\begin{equation*}
\boldsymbol{y} = \nabla \psi (\boldsymbol{z}) \sim P.
\end{equation*}
From Rockafellar's Theorem \citep{rockafellar_characterization_1966} we conclude: 1. that $\nabla \psi$ must be cyclically monotone, 2. if there exists a cyclically monotone function $G$ pushing forward the standard normal distribution on $\mathbb{R}^p$ to $P$, then the graph of $G$ must coincide with the graph of the gradient of a convex function; hence, by the uniqueness above we must have $G = \nabla \psi$ almost-everywhere. Therefore, there is an almost-everywhere unique $G \in \mathcal{G}^p$ so that $\boldsymbol{y} = G(\boldsymbol{z}) \sim P$.
\end{proof}

This proposition shows that there exist unique cyclically monotone transformation pairs $G_1$, $G_2$ that yield arbitrary marginal distributions $P_1, P_2$ for the rows of the data matrices $\boldsymbol Y_1$, $\boldsymbol Y_2$. Additionally, in the submodel for which the multivariate margins $P_1$ and $P_2$ are absolutely continuous with respect to Lebesgue measure, our model for semiparametric CCA is identifiable up to simultaneous permutation and sign change of the columns of $\boldsymbol{Q}_1, \boldsymbol{Q}_2$, an ambiguity arising due to the matrix product of the CCA parameters.
\begin{proposition}\label{prop:p2}
    Let $\boldsymbol{\theta} = (G_1, G_2, \boldsymbol{Q}_1, \boldsymbol{Q}_2, \boldsymbol{\Lambda})$ and let $P_{\boldsymbol{\theta}}$ denote the joint probability distribution of an observation $(\boldsymbol{y}_{i,1}, \boldsymbol{y}_{i,2})$ from the model (\ref{eq:cmcca_mod}) with absolutely continuous $p_1$- and $p_2$-dimensional marginal distributions $P_1, P_2$. Then if $P_{\boldsymbol{\theta}} = P_{\boldsymbol{\theta}'}$, the following hold:
\begin{enumerate}[noitemsep]
    \item $G_1 = G'_1,~G_2 = G'_2$;
    \item $\boldsymbol{\Lambda} = \boldsymbol{\Lambda}'$;
    \item $\boldsymbol{Q}_1 = \boldsymbol{Q}'_1 \boldsymbol{P} \boldsymbol{S}, \boldsymbol{Q}_2 = \boldsymbol{Q}'_2 \boldsymbol{P} \boldsymbol{S}$, where $\boldsymbol{S} \in \mathbb{R}^{d \times d}$ is any diagonal matrix with diagonal entries in $\left\{-1, 1 \right\}$ and $\boldsymbol{P}$ is any permutation matrix with $\boldsymbol{P}_k = \boldsymbol{e}_k$ if $\lambda_k \neq \lambda_l~\forall l \neq k$.
\end{enumerate}
\end{proposition}

\begin{proof}[Proof of Proposition 2]

Suppose that
\begin{equation*}
P_{\boldsymbol{\theta}} = P_{\boldsymbol{\theta}'}
\end{equation*}
The equality of the joint probability distributions above implies that the corresponding $p_1$- and $p_2$- dimensional margins must also be equal. As before, we apply McCann's Theorem \citep{mccann_existence_1995} and Rockafellar's Theorem \citep{rockafellar_characterization_1966} to find that the transformations $G_1, G_2$ pushing forward the $p_1$- and $p_2$- dimensional standard normal distributions to $P_1$, $P_2$, respectively, must be unique almost everywhere. Thus,
\begin{equation*}
G_1 = G_1',~~G_2 = G_2'~~~a.e.
\end{equation*}
Assuming that $P_1, P_2$ are absolutely continuous, McCann's Theorem \citep{mccann_existence_1995} and Rockafellar's Theorem \citep{rockafellar_characterization_1966} also guarantee the existence and almost-everywhere uniqueness of cyclically monotone functions $H_1, H_2$ pushing forward $P_1$, $P_2$ to the $p_1$- and $p_2$-dimensional standard normal distributions, respectively. Moreover, $H_1 \circ G_1(\boldsymbol{z}_1) = \boldsymbol{z}_1$ a.e. and $H_2 \circ G_2(\boldsymbol{z}_2) = \boldsymbol{z}_2$ a.e. Applying these functions to $\boldsymbol{y}_{1}$ and $\boldsymbol{y}_{2}$, we obtain 
\begin{equation*}
\left[\begin{array}{c}
H_1(\boldsymbol{y}_1) \\
H_2(\boldsymbol{y}_2)
\end{array}\right] \sim N_{p_1 + p_2} \left( \boldsymbol{0},  \left[ \begin{array}{cc}
\boldsymbol{I}_{p_1} & \boldsymbol{Q}_1 \boldsymbol{\Lambda} \boldsymbol{Q}_2^\top \\
\boldsymbol{Q}_2 \boldsymbol{\Lambda} \boldsymbol{Q}_1^\top & \boldsymbol{I}_{p_2} \\
\end{array} \right]\right),
\end{equation*}
and, likewise,
\begin{equation*}
\left[\begin{array}{c}
H_1(\boldsymbol{y}_1') \\
H_2(\boldsymbol{y}_2')
\end{array}\right] \sim N_{p_1 + p_2} \left( \boldsymbol{0},  \left[ \begin{array}{cc}
\boldsymbol{I}_{p_1} & \boldsymbol{Q}_1' \boldsymbol{\Lambda}' {\boldsymbol{Q}'_2}^\top \\
\boldsymbol{Q}_2' \boldsymbol{\Lambda}' {\boldsymbol{Q}_1'}^\top & \boldsymbol{I}_{p_2} \\
\end{array} \right]\right),
\end{equation*}
Because $(H_1(\boldsymbol{y}_1), H_2(\boldsymbol{y}_2)) \overset{d}{=} (H_1(\boldsymbol{y}_1'), H_2(\boldsymbol{y}_2'))$, we obtain
\begin{equation*}
\boldsymbol{Q}_1 \boldsymbol{\Lambda} {\boldsymbol{Q}_2}^\top = \boldsymbol{Q}_1' \boldsymbol{\Lambda}' {\boldsymbol{Q}'_2}^\top
\end{equation*}
If the diagonal elements of $\boldsymbol{\Lambda}, \boldsymbol{\Lambda}'$  are strictly decreasing values $[0, 1)$, then the uniqueness of the matrix singular value decomposition yields
\begin{equation*}
\boldsymbol{\Lambda} = \boldsymbol{\Lambda}'
\end{equation*}
and
\begin{equation*}
\boldsymbol{Q}_1 = \boldsymbol{Q}'_1 \boldsymbol{S}, \boldsymbol{Q}_2 = \boldsymbol{Q}'_2 \boldsymbol{S},
\end{equation*}
where $\boldsymbol{S} \in \mathbb{R}^{d \times d}$ is an arbitrary diagonal matrix with diagonal entries in $\left\{-1, 1\right\}$. If the entries of $\boldsymbol{\Lambda}, \boldsymbol{\Lambda}'$ are not strictly decreasing, then the uniqueness of the matrix singular value decomposition yields
\begin{equation*}
\boldsymbol{\Lambda} = \boldsymbol{\Lambda}'
\end{equation*}
and 
\begin{equation*}
\boldsymbol{Q}_1 = \boldsymbol{Q}'_1 \boldsymbol{P} \boldsymbol{S}, \boldsymbol{Q}_2 = \boldsymbol{Q}'_2 \boldsymbol{P} \boldsymbol{S},
\end{equation*}
where $\boldsymbol{P}$ is any permutation matrix with $\boldsymbol{P}_k = \boldsymbol{e}_k$ if $\lambda_k \neq \lambda_l~\forall l \neq k$ and $\boldsymbol{S}$ is as before.
\end{proof}

The preceding two propositions show that our model for semiparametric CCA using cyclically monotone transformations is not only an extension of the model for traditional CCA, but also an extension of models used in existing methods for semiparametric CCA \citep{zoh_pcan_2016, agniel_analysis_2017, yoon_sparse_2020}. These methods infer the CCA parameters indirectly through inference of a $(p_1 + p_2) \times (p_1 + p_2)$ correlation matrix, which parameterizes a $(p_1 + p_2)$-dimensional Gaussian copula model. While the Gaussian copula model is semiparametric in the sense that the univariate margins of the variable sets may be arbitrary, it still assumes that the multivariate marginal distributions of the variable sets (and any subsets of those variables) are distributed according to a Gaussian copula model. By contrast, our model for semiparametric CCA using cyclically monotone transformations allows for arbitrary multivariate marginal distributions, even those that cannot be described by a Gaussian copula. While the identifiability result of Proposition \ref{prop:p2} is stated with respect to the submodel in which the multivariate margins $P_1, P_2$ are absolutely continuous, an informal argument suggests that identifiability of the CCA parameters should also hold in the case of multivariate margins with at least one continuous part. For instance, if the marginal distribution $P_1$ is absolutely continuous on a subset $A \subseteq \mathbb{R}^{p_1}$, then $G_1$ must be invertible on $A$, so that the density of $\boldsymbol{z}_{i,1}$ is identified on a set of positive measure in $\mathbb{R}^{p_1}$. A similar argument applied to the second margin implies that the joint density of $(\boldsymbol{z}_{i,1}, \boldsymbol{z}_{i,2})$ can be identified on a set of positive measure in $\mathbb{R}^{p_1 + p_2}$. Since the multivariate normal density function is analytic, this identifies the density everywhere, which identifies the CCA parameters up to the ambiguities mentioned in Proposition \ref{prop:p2}.

Before discussing estimation and inference strategies for the CCA parameters, we conclude this section with some additional intuition for cyclically monotone functions. Cyclical monotonicity is a property of subsets of $\mathbb{R}^p \times \mathbb{R}^p$ that generalizes one-dimensional monotonicity to $p \geq 2$, and cyclically monotone functions coincide exactly with the set of gradients of convex functions \citep{rockafellar_characterization_1966}, so they possess a regularity that is similar to that of positive definite, symmetric linear operators. In fact, the linear function $\mathbf{z} \rightarrow \boldsymbol{\Sigma}^{1/2} \mathbf{z}$ is cyclically monotone for $\boldsymbol{\Sigma}$ symmetric and positive definite, so the semiparametric CCA model 
(\ref{eq:cmcca_mod}) generalizes the Gaussian CCA model (\ref{eq:norm_mod}). Other examples of cyclically monotone functions include coordinate-wise monotone functions, radially symmetric functions that scale vectors by monotonic functions of their norms (e.g. $\mathbf{z} \rightarrow \mathbf{z} / \|\mathbf{z}\|$), and functions of the form $\mathbf{z} \rightarrow \boldsymbol{B}^\top G ( \boldsymbol{B} \mathbf{z})$ where $\boldsymbol{B}$ is a matrix and $G$ is a cyclically monotone function. Cyclically monotone functions also correspond to optimal transport maps between measures with finite second moments \citep{brenier_polar_1991, ambrosio_users_2013} when such maps exist.

\section{Inference for the CCA parameters}\label{sec:3}
For the moment, consider the submodel in which $G_1$ and $G_2$ 
are restricted so that the marginal distributions they induce have
densities with respect to Lebesgue measure. Let 
$\mathbf{Y} \in \mathbb R^{n\times (p_1+p_2)}$ be the observed value of 
$\boldsymbol{Y}$
and denote the corresponding likelihood as
\begin{equation}
    L(\boldsymbol{Q}_1, \boldsymbol{Q}_2, \boldsymbol{\Lambda}, G_1, G_2 : \mathbf{Y}) = p\left(\mathbf{Y} | \boldsymbol{Q}_1, \boldsymbol{Q}_2, \boldsymbol{\Lambda}, G_1, G_2 \right) .
\end{equation}
Inference for semiparametric CCA is challenging because the likelihood depends on the infinite dimensional parameters $G_1, G_2$. If $G_1, G_2$ were known and invertible, inference for the CCA parameters could proceed by maximizing a multivariate Gaussian likelihood in $\boldsymbol{Q}_1, \boldsymbol{Q}_2, \boldsymbol{\Lambda}$ using $\boldsymbol{z}_{i,j} = G_j^{-1}(\boldsymbol{y}_{i,j})$ as the data. However, in practice the $G_j$'s are unknown. 

\subsection{Inference using pseudolikelihoods}

Faced with similar problems in semiparametric copula estimation---for which the marginal transformations are univariate monotone functions $g_j$---previous authors have used pseudolikelihood methods \citep{oakes_multivariate_1994}. One pseudolikelihood method is to construct an estimate $\hat{g}_j$ for each univariate transformation function and maximize the parametric copula likelihood using $\hat{z}_{i,j} = \hat{g}_j^{-1} (y_{i,j})$ as a plug-in estimate for the data. Under certain conditions, this procedure has been shown to produce consistent and asymptotically normal estimates \citep{genest_semiparametric_1995}. Such an approach can be implemented for estimation of the CCA parameters, using score functions of the multivariate ranks of \cite{hallin_distribution_2017} or \cite{deb_multivariate_2021} as plug-in estimates for the latent $\boldsymbol{z}_{i,j}$'s in (\ref{eq:cmcca_mod}). Given plug-in data $\hat{\boldsymbol{Z}}_1, \hat{\boldsymbol{Z}}_2$, reasonable estimates for the CCA parameters might be the singular vectors and singular values of the matrix $(\tilde{\boldsymbol{Z}}_1^\top \tilde{\boldsymbol{Z}}_1)^{-1/2} \tilde{\boldsymbol{Z}}_1^\top \tilde{\boldsymbol{Z}}_2 (\tilde{\boldsymbol{Z}}_2^\top \tilde{\boldsymbol{Z}}_2)^{-1/2}$ where $\tilde{\boldsymbol{Z}}_1 = \hat{\boldsymbol{Z}}_1 - \frac{1}{n} \boldsymbol{1} \boldsymbol{1}^\top \hat{\boldsymbol{Z}}_1$, just as in traditional CCA. In the case that $G_1, G_2$ are continuous (see \cite{figalli_continuity_2018, del_barrio_note_2020} for continuity conditions when the reference measure is spherical uniform on the unit ball) the consistency of such estimates for the CCA parameters follows from Proposition 5.1 of \cite{hallin_distribution_2017} combined with an application of the continuous mapping theorem. Confidence intervals for the CCA parameters might then be obtained using the asymptotic normal approximations of \cite{anderson_asymptotic_1999}.  See Section \ref{par:pseudo} for details on a specific method for computing pseudolikelihood estimates for the CCA parameters.

This plug-in approach is similar to those found in the literature for constructing statistics used in non-parametric tests for independence and equality in distribution \citep{deb_efficiency_2021, deb_multivariate_2021} and should work well for estimation tasks with large sample sizes. In the next section, we derive a likelihood function leading to a Bayesian approach to inference of the CCA parameters, which does not depend on asymptotic arguments, handles data missing at random, and simplifies uncertainty quantification for arbitrary functions of the CCA parameters. 

\subsection{Bayesian inference using the multirank likelihood} 

Another approach to semiparametric 
copula inference is to use a type of marginal likelihood based 
on the univariate ranks of each variable, called the rank 
likelihood \citep{hoff_extending_2007, hoff_information_2014}. The rank likelihood
is a function of the copula parameters only, and provides inferences
that are invariant to strictly monotonic transformations of each variable. Here we generalize this approach to the 
semiparametric CCA model, by constructing what we call a 
\textit{multirank likelihood}, which can be interpreted as a likelihood 
for the CCA parameters based on information from the data that 
does not depend on the infinite-dimensional parameters 
$G_1$ and $G_2$. This information can be characterized as follows: 
because each $\boldsymbol{y}_{i,j}$ is a cyclically monotone transformation of the latent variable $\boldsymbol{z}_{i,j}$, the matrices $\boldsymbol{Z}_j$ and $\boldsymbol{Y}_j$ must be in \textit{cyclically monotone correspondence}, meaning $\boldsymbol{Z}_j$ has to lie in the set
\begin{equation}
    \mathcal{K}(\boldsymbol{Y}) := \left\{ \boldsymbol{Z}_j \in \mathbb{R}^{n \times p_j} : \{(\boldsymbol{z}_{i, j}, \boldsymbol{y}_{i, j})\}_{i=1}^n \text{ is cyclically monotone}\right\}.
\end{equation}
Let $\boldsymbol{Z} \in \mathcal{K}(\boldsymbol{Y})$ denote that both $\boldsymbol{Z}_1, \boldsymbol{Y}_1$ and $\boldsymbol{Z}_2, \boldsymbol{Y}_2$ are in cyclically monotone correspondence, 
and note that under our model, $\boldsymbol{Z} \in \mathcal{K}(\boldsymbol{Y})$ with probability 1. 
Now suppose it is observed that $\boldsymbol{Y} = \mathbf{Y}$ 
for some $\mathbf{Y} \in \mathbb R^{n\times (p_1+p_2)}$. Then part of the information from the data is that $\boldsymbol{Z} \in \mathcal{K}(\mathbf{Y})$. 
We define the multirank likelihood $L_{\rm M}$ to be the probability of this event, as a function of the model parameters:
\begin{align*}
\Pr( \boldsymbol{Z} \in \mathcal{K}(\mathbf{Y}) | \boldsymbol{Q}_1, \boldsymbol{Q}_2, \boldsymbol{\Lambda}, G_1, G_2) 
& = \Pr( \boldsymbol{Z} \in \mathcal{K}(\mathbf{Y}) | \boldsymbol{Q}_1, \boldsymbol{Q}_2, \boldsymbol{\Lambda}) \\
 & \equiv L_{\rm M}( \boldsymbol{Q}_1, \boldsymbol{Q}_2, \boldsymbol{\Lambda}  : \mathbf{Y}). 
\end{align*} 
The multirank likelihood depends on the CCA parameters only
and not on the infinite-dimensional 
parameters $G_1$ and $G_2$ because the distribution of $\boldsymbol{Z}$ does not depend on $G_1,G_2$. 

The multirank likelihood can be interpreted as a type of marginal likelihood as follows: for convenience, consider the case where the margins 
of $\boldsymbol{Y}$ are discrete. Having observed 
$\boldsymbol{Y} = \mathbf{Y}$ for some $\mathbf{Y}\in \mathbb R^{n\times (p_1+p_2)}$, 
the full likelihood in terms of $\boldsymbol{\theta} = (\boldsymbol{Q}_1, \boldsymbol{Q}_2, \boldsymbol{\Lambda})$ can be decomposed as follows:
\begin{align}
L(\boldsymbol{\theta}, G_1,G_2 :\mathbf{Y} ) &= 
\Pr(\boldsymbol{Y}=\mathbf{Y} | \boldsymbol{\theta} , G_1,G_2 ) \nonumber \\
&= \Pr(\boldsymbol{Z} \in \mathcal{K}(\boldsymbol{Y}), 
\boldsymbol{Y}= \mathbf{Y}  | \boldsymbol{\theta} , G_1,G_2 ) \nonumber \\
&=  \Pr(\boldsymbol{Z} \in \mathcal{K}(\mathbf{Y}), 
\boldsymbol{Y}=\mathbf{Y}  | \boldsymbol{\theta} , G_1,G_2 ) \nonumber \\
&=  \Pr(\boldsymbol{Z} \in \mathcal{K}(\mathbf{Y}) | 
    \boldsymbol{\theta}) \times \Pr(\boldsymbol{Y}=\mathbf{Y}  | \boldsymbol{Z} \in \mathcal{K}(\mathbf{Y}) ,\boldsymbol{\theta} , G_1,G_2 ) \nonumber \\
& = L_{\rm M}(\boldsymbol{\theta} :\mathbf{Y}) \times \Pr(\boldsymbol{Y}=\mathbf{Y}  | \boldsymbol{Z} \in \mathcal{K}(\mathbf{Y}) ,\boldsymbol{\theta}, G_1,G_2 ), \nonumber
\end{align}
where the second line holds because 
$\boldsymbol{Z} \in \mathcal{K}(\boldsymbol{Y})$ 
happens with probability one, and the third line substitutes the random $\boldsymbol{Y}$ for the observed value $\mathbf{Y}$. Thus the multirank likelihood 
can be viewed as a type of marginal likelihood \cite[Section 8.3]{severini_likelihood_2000}, derived from the marginal probability of the event $\boldsymbol{Z} \in \mathcal{K}(\mathbf{Y})$. 

Letting $\Gamma_{\text{CCA}}$ denote the probability density function for the matrix normal distribution with the CCA parameterization \eqref{eq:norm_mod}, we see that applying the maximum likelihood principle to the multirank likelihood is infeasible because of the intractable integral $\int_{\mathcal{K}(\mathbf{Y})} \Gamma_{\text{CCA}}(\mathbf{Z} | \boldsymbol{Q}_1, \boldsymbol{Q}_2, \boldsymbol{\Lambda}) d \mathbf{Z}$. By contrast, Bayesian inference for the CCA parameters using the multirank likelihood is tractable. Specifically, summaries of the posterior distribution of the CCA parameters, written as
\begin{equation*}
    p(\boldsymbol{Q}_1, \boldsymbol{Q}_2, \boldsymbol{\Lambda} | \boldsymbol{Z} \in \mathcal{K}(\mathbf{Y})) \propto  L_M(\boldsymbol{Q}_1, \boldsymbol{Q}_2, \boldsymbol{\Lambda} : \mathbf{Y}) \times p(\boldsymbol{Q}_1, \boldsymbol{Q}_2, \boldsymbol{\Lambda}),
\end{equation*}
can be approximated by constructing a Markov Chain with stationary distribution equal to this posterior. Values simulated from this Markov Chain can then be used to obtain estimates and confidence regions for the CCA parameters. The probabilities described by these confidence regions are Bayesian posterior probabilities based on the partial information in $L_{\rm M}(\boldsymbol{\theta} :\mathbf{Y})$, rather than the full information in $L(\boldsymbol{\theta}, G_1,G_2 :\mathbf{Y} )$ (see e.g. \cite{severini_likelihood_2000} or \cite{pauli_bayesian_2011} for reviews of Bayesian inference in the context of partial, marginal, and composite likelihoods).

We construct the desired Markov Chain using a data augmentation scheme \citep{dempster_maximum_1977, gelfand_sampling-based_1990}, in which a Markov Chain is iterated over the product space of the latent variables $\boldsymbol{Z}_1, \boldsymbol{Z}_2$ and the CCA parameters. Iterates of the Markov Chain consist of values of each of the CCA parameters and each of the latent variables $\boldsymbol{Z}_1, \boldsymbol{Z}_2$ simulated using a Metropolis-within-Gibbs algorithm. In each sub-step of the algorithm, a subset of the variables are held constant, and a Metropolis-Hastings step is used to simulate the remaining variables using their full conditional distribution as the target. The full conditional distribution of each variable subset is proportional to $\Gamma_{\text{CCA}}(\boldsymbol{Z} | \boldsymbol{Q}_1, \boldsymbol{Q}_2, \boldsymbol{\Lambda}) \mathbf{1}_{\boldsymbol{Z} \in \mathcal{K}(\mathbf{Y})}$ times the prior distribution on the CCA parameters. Since the supports of the CCA parameters are, respectively, the Stiefel manifolds $\mathcal{V}_{p_1, d}$, $\mathcal{V}_{p_2, d}$, and the subset of $[0, 1)^d$ on which $\lambda_1 \geq \dots \geq \lambda_d$, we choose to set independent, uniform priors on these compact spaces. The simulation steps are then as follows:

\paragraph{Simulating $\boldsymbol{Q}_1, \boldsymbol{Q}_2$}

With a uniform prior on $\mathcal{V}_{p_1, d}$, the full conditional distribution of $\boldsymbol{Q}_1$ is given by
\begin{align}\label{eq:qsim}
    \boldsymbol{Q}_1 | - &\sim \text{BMF}(\boldsymbol{A}_1, \boldsymbol{B}_1, \boldsymbol{C}_1) \\
    &\boldsymbol{A}_1 = -\boldsymbol{Z}_1^\top \boldsymbol{Z}_1 / 2, \boldsymbol{B}_1 = (\boldsymbol{\Lambda}^{-2} - \boldsymbol{I})^{-1}, \boldsymbol{C}_1 = \boldsymbol{Z}_1^\top \boldsymbol{Z}_2 \boldsymbol{Q}_2 (\boldsymbol{\Lambda}^{-1} - \boldsymbol{\Lambda})^{-1}, \nonumber
\end{align}
where $\mathrm{BMF}$ denotes the Bingham-von Mises-Fisher distribution parameterized as in \cite{hoff_simulation_2009}. To simulate new values of $\boldsymbol{Q}_1$ with this as the target distribution, we combine elliptical slice sampling based on matrix-normal priors \citep{murray_elliptical_2010} with the polar expansion strategy of \cite{jauch_monte_2021}. This involves iterating a Markov Chain for an auxiliary variable $\boldsymbol{X}_1$ and taking $\boldsymbol{Q}_1$ to be the left polar factor of $\boldsymbol{X}_1$ at each iteration. Specifically, we propose a new value for $\boldsymbol{Q}_1$ by first proposing $\boldsymbol{X}_1^*$, and then setting 
\begin{equation*}
    \boldsymbol{Q}_1^* = \boldsymbol{U}_1^* \boldsymbol{V}_1^{*\top},
\end{equation*}
where $\boldsymbol{U}_1^*, \boldsymbol{V}_1^*$ are the left and right singular vectors of $\boldsymbol{X}_1^*$. Proposals $\boldsymbol{X}_1^*$ are generated using the elliptical slice sampling algorithm, which is valid when targeting a density proportional to the product of a likelihood function and a normal prior density. Targeting the BMF density \eqref{eq:qsim} as a function of $\boldsymbol{X}_1$ times the density for the normal prior distribution $\boldsymbol{X}_1 \sim N_{p_1 \times d}(\boldsymbol{0}, \boldsymbol{I}_d \otimes \boldsymbol{I}_{p_1})$ yields the desired stationary distribution for $\boldsymbol{Q}_1$ after transformation of variables \citep{eaton_multivariate_1983}. To simulate $\boldsymbol{Q}_2$, we apply the same steps with the variable subscripts interchanged.

\paragraph{Simulating $\boldsymbol{\Lambda}$}

The full conditional distribution for $\boldsymbol{\Lambda}$ has density proportional to
\begin{align}
    &\prod_{k=1}^d (1 - \lambda_k^2)^{-n/2} e^{-\frac{((a_k + b_k)/2 + c_k)}{2(1 + \lambda_k)}} e^{-\frac{((a_k + b_k)/2 - c_k)}{2(1 - \lambda_k)}} \mathbf{1}_{1 > \lambda_1 \geq \dots \geq \lambda_d \geq 0}\label{lambda_fc} \\
    &\boldsymbol{a} = \mathrm{diag}(\boldsymbol{Q}_1^\top \boldsymbol{Z}_1^\top \boldsymbol{Z}_1 \boldsymbol{Q}_1), \boldsymbol{b} = \mathrm{diag}(\boldsymbol{Q}_2^\top \boldsymbol{Z}_2^\top \boldsymbol{Z}_2 \boldsymbol{Q}_2), \boldsymbol{c} = \mathrm{diag}(\boldsymbol{Q}_1^\top \boldsymbol{Z}_1^\top \boldsymbol{Z}_2 \boldsymbol{Q}_2). \nonumber
\end{align}
This density has an exponential family form, but it is not one for which standard simulation routines are available. We therefore choose to use a Metropolis-Hastings step to simulate each $\lambda_k$, proposing new values according to
\begin{equation*}
\lambda_k^* \sim N\left(r_k, \frac{1}{n}\left(1 - \frac{2 c_k}{a_k + b_k}\right) \right),
\end{equation*}
truncated to the region $\lambda_{k-1} \geq \lambda_k^* \geq \lambda_{k+1}$, where $r_k$ is the real root to the cubic equation
\begin{equation*}
\lambda_k^3 - \frac{c_k}{n} \lambda_k^2 + \frac{a_k+b_k-n}{n} \lambda_k - \frac{c_k}{n} = 0,
\end{equation*}
This equation is obtained by setting the derivative of the logarithm of the full conditional density for $\lambda_k$ to zero, so that $r_k$ is the mode of the full conditional distribution \eqref{lambda_fc}.

\paragraph{Simulating $\boldsymbol{Z}_1, \boldsymbol{Z}_2$}

The full conditional distribution of $\boldsymbol{Z}_1$ is
\begin{align}
    \boldsymbol{Z}_1 | - &\sim N^{\mathcal{K}(\mathbf{Y}_1)}_{n \times p_1}(\boldsymbol{B}_1, \boldsymbol{A}_1 \otimes \boldsymbol{I}_n)\\
    &\boldsymbol{A}_1 = \boldsymbol{I}_{p_1} - \boldsymbol{Q}_1 \boldsymbol{\Lambda}^2 \boldsymbol{Q}_1^\top, \boldsymbol{B}_1 = \boldsymbol{Z}_2 \boldsymbol{Q}_2 \boldsymbol{\Lambda} \boldsymbol{Q}_1^\top, \nonumber
\end{align}
where $N^{ \mathcal{K}(\mathbf{Y}_1)}_{n \times p_1}$ refers to the matrix normal distribution truncated to the set of matrices in cyclically monotone correspondence with $\mathbf{Y}_1$. Simulating values of $\boldsymbol{Z}_1$ constrained to lie in $\mathcal{K}(\mathbf{Y}_1)$ presents several challenges, since directly checking the condition in Definition \ref{def:cm} requires a number of operations that scales combinatorially with $n$. However, the equivalent condition
\begin{equation}\label{eq:optim}
    \sum_{i=1}^n \|\mathbf{z}_i - \mathbf{y}_i\|_2^2 = \underset{\sigma = (\sigma(1), \dots, \sigma(n)) \in S_n}{\min} \sum_{i=1}^n \|\mathbf{z}_i - \mathbf{y}_{\sigma(i)}\|_2^2.
\end{equation}
can be checked in $O(n^3)$ time using algorithms for solving the optimal assignment problem between the point sets $\{\mathbf{z}_1, \dots, \mathbf{z}_n\}$ and $\{\mathbf{y}_1, \dots, \mathbf{y}_n\}$ \citep{peyre_computational_2018}. Our approach to simulating values of $\boldsymbol{Z}_1$ relies on using the dual to the assignment problem to check optimality after updates are proposed in a Metropolis-Hastings step. Within each step, we use a weaker condition than cyclical monotonicity, which is sometimes called \textit{2-monotonicity}, or simply \textit{monotonicity}, to inform our proposal distribution for $\boldsymbol{z}_{i, 1}$. Specifically, if $\boldsymbol{z}_{i, 1}^{(t)}$ is an MCMC iterate for $\boldsymbol{z}_{i, 1}$ at iteration $t$, then we propose a new iterate according to
\begin{equation}\label{eq:prop}
    \boldsymbol{z}_{i, 1}^* \sim N_{p_1}( \boldsymbol{z}_{i, 1}^{(t)}, \boldsymbol{A}^{-1} / \chi_{0.95}^2),
\end{equation}
where
\begin{equation*}
    \boldsymbol{A} = \sum_{j \neq i} \frac{1}{\{(\boldsymbol{z}_{i, 1}^{(t)} - \boldsymbol{z}_{j, 1}^{(t)})^\top (\mathbf{y}_{i, 1} - \mathbf{y}_{j, 1})\}^2} (\mathbf{y}_{i, 1} - \mathbf{y}_{j, 1})(\mathbf{y}_{i, 1} - \mathbf{y}_{j, 1})^\top,
\end{equation*}
and $\chi_{0.95}^2$ denotes the $0.95$ quantile of the $\chi^2$ distribution with $p_1$ degrees of freedom. If $\boldsymbol{Z}_1^*$---the matrix obtained by setting the $i^\text{th}$ row of $\boldsymbol{Z}_1^{(t)}$ to $\boldsymbol{z}_{i, 1}^*$---is in $\mathcal{K}(\mathbf{Y}_1)$, then we accept $\boldsymbol{z}_{i, 1}^*$ with a Metropolis-adjusted probability.

This particular choice of proposal distribution is justified by the definition of monotonicity.
\begin{definition}[Monotonicity]\label{def:mono}
    A subset $S$ of $\mathbb{R}^p \times \mathbb{R}^p$ is \textit{monotone} if and only if
    \begin{equation}\label{eq:mono}
        (\mathbf{z}_i - \mathbf{z}_j)^\top (\mathbf{y}_i - \mathbf{y}_j) \geq 0
    \end{equation}
    for any finite collection of points $\left\{ (\mathbf{z}_1, \mathbf{y}_1), \dots, (\mathbf{z}_n, \mathbf{y}_n) \right\} \subseteq S$ and all $i, j \in \{1, \dots, n\}$.
\end{definition}
This condition implies that the probability of preserving monotonicity with the proposed update can be bounded from below, as shown by the following result.
\begin{proposition}
    Let $\boldsymbol{z}_{i, 1}^*$ be distributed as in \eqref{eq:prop} and assume that the set $\{(\boldsymbol{z}_{i, 1}^{(t)}, \mathbf{y}_{i, 1})\}_{i=1}^n$ is monotone with strict inequalities in the conditions of \eqref{eq:mono}. Then the set obtained by replacing $\boldsymbol{z}_{i, 1}^{(t)}$ with $\boldsymbol{z}_{i, 1}^*$ is monotone with probability at least $0.95$.
\end{proposition}
\begin{proof}[Proof of Proposition 3]
If $\boldsymbol{z}_{i,1}^*$ is distributed with mean and covariance as in \eqref{eq:prop}, then $(\boldsymbol{z}_{i,1}^* - \boldsymbol{z}_{i,1}^{(t)})^\top (\chi_{0.95}^2 \boldsymbol{A})(\boldsymbol{z}_{i,1}^* - \boldsymbol{z}_{i,1}^{(t)})$ is distributed as a chi-square random variable with $p_1$ degrees of freedom. Therefore,
\begin{equation*}
    \text{Pr}((\boldsymbol{z}_{i,1}^* - \boldsymbol{z}_{i,1}^{(t)}) \boldsymbol{A} (\boldsymbol{z}_{i,1}^* - \boldsymbol{z}_{i,1}^{(t)}) < 1) = 0.95
\end{equation*}
The event $(\boldsymbol{z}_{i,1}^* - \boldsymbol{z}_{i,1}^{(t)}) \boldsymbol{A} (\boldsymbol{z}_{i,1}^* - \boldsymbol{z}_{i,1}^{(t)}) < 1$ implies
\begin{equation*}
    \begin{aligned}
        \sum_{j \neq i} \frac{\{(\boldsymbol{z}_{i,1}^* - \boldsymbol{z}_{i,1}^{(t)})^\top (\mathbf{y}_{i, 1} - \mathbf{y}_{j, 1})\}^2}{\{(\boldsymbol{z}_{i, 1}^{(t)} - \boldsymbol{z}_{j, 1}^{(t)})^\top (\mathbf{y}_{i, 1} - \mathbf{y}_{j, 1})\}^2} < 1  &\implies \\
        \frac{\{(\boldsymbol{z}_{i,1}^* - \boldsymbol{z}_{i,1}^{(t)})^\top (\mathbf{y}_{i, 1} - \mathbf{y}_{j, 1})\}^2}{\{(\boldsymbol{z}_{i, 1}^{(t)} - \boldsymbol{z}_{j, 1}^{(t)})^\top (\mathbf{y}_{i, 1} - \mathbf{y}_{j, 1})\}^2} < 1,~\forall j \neq i &\implies \\
        \left\{\frac{(\boldsymbol{z}_{i,1}^* - \boldsymbol{z}_{j, 1}^{(t)})^\top(\mathbf{y}_{i, 1} - \mathbf{y}_{j, 1}) - (\boldsymbol{z}_{i,1}^{(t)} - \boldsymbol{z}_{j, 1}^{(t)})^\top (\mathbf{y}_{i, 1} - \mathbf{y}_{j, 1})}{(\boldsymbol{z}_{i, 1}^{(t)} - \boldsymbol{z}_{j, 1}^{(t)})^\top (\mathbf{y}_{i, 1} - \mathbf{y}_{j, 1})}\right\}^2 < 1,~\forall j \neq i &\implies \\
        \left|\frac{(\boldsymbol{z}_{i,1}^* - \boldsymbol{z}_{j, 1}^{(t)})^\top(\mathbf{y}_{i, 1} - \mathbf{y}_{j, 1})}{(\boldsymbol{z}_{i, 1}^{(t)} - \boldsymbol{z}_{j, 1}^{(t)})^\top (\mathbf{y}_{i, 1} - \mathbf{y}_{j, 1})} - 1\right| < 1,~\forall j \neq i &\implies \\
        0 < \frac{(\boldsymbol{z}_{i,1}^* - \boldsymbol{z}_{j, 1}^{(t)})^\top(\mathbf{y}_{i, 1} - \mathbf{y}_{j, 1})}{(\boldsymbol{z}_{i, 1}^{(t)} - \boldsymbol{z}_{j, 1}^{(t)})^\top (\mathbf{y}_{i, 1} - \mathbf{y}_{j, 1})} < 2,~\forall j \neq i
    \end{aligned}
\end{equation*}
Since $\{(\boldsymbol{z}_{i, 1}^{(t)}, \mathbf{y}_{i, 1})\}_{i=1}^n$ is strictly monotone by assumption, the denominators in the last line are positive for all $j \neq i$. Hence, the numerators are also positive, and we have
\begin{equation*}
    (\boldsymbol{z}_{i,1}^* - \boldsymbol{z}_{i,1}^{(t)}) \boldsymbol{A} (\boldsymbol{z}_{i,1}^* - \boldsymbol{z}_{i,1}^{(t)}) < 1 \implies (\boldsymbol{z}_{i,1}^* - \boldsymbol{z}_{j, 1}^{(t)})^\top(\mathbf{y}_{i, 1} - \mathbf{y}_{j, 1}) > 0,~\forall i \neq j,
\end{equation*}
which is the monotonicity condition. Therefore, the set obtained by replacing $\boldsymbol{z}_{i, 1}^{(t)}$ with $\boldsymbol{z}_{i, 1}^*$ is monotone with probability at least $\text{Pr}((\boldsymbol{z}_{i,1}^* - \boldsymbol{z}_{i,1}^{(t)}) \boldsymbol{A} (\boldsymbol{z}_{i,1}^* - \boldsymbol{z}_{i,1}^{(t)}) < 1) = 0.95$.
\end{proof}

Since monotonicity is necessary, but not sufficient to guarantee cyclical monotonicity, the result above, as well as our choice of proposal distribution, are merely heuristics used to increase the rate of acceptance within each Metropolis-Hastings step. To verify the cyclical monotonicity of a proposed update, we turn to the formulation of the assignment problem as a linear program. The complementary slackness conditions of this program imply that \eqref{eq:optim} holds if and only if there exist real vectors $\boldsymbol{u}, \boldsymbol{v} \in \mathbb{R}^n$, sometimes called \textit{dual variables}, such that
\begin{equation*}
    \begin{aligned}
        u_i + v_i &= \|\mathbf{z}_i - \mathbf{y}_{i}\|_2^2,~i=1,\dots, n \\
        u_i + v_j &\leq \|\mathbf{z}_{i} - \mathbf{y}_{j}\|_2^2,~i\neq j.
    \end{aligned}
\end{equation*}
Making substitutions using the equality constraints above and collecting terms leads to the following necessary and sufficient condition for cyclical monotonicity: a proposed matrix $\boldsymbol{Z}_1^*$ is cyclically monotone with $\mathbf{Y}_1$ if and only if there exists a $\boldsymbol{w} \in \mathbb{R}^n$ such that
\begin{equation}\label{eq:slack_cond}
    \boldsymbol{w} \mathbf{1}_n^\top - \mathbf{1}_n \boldsymbol{w}^\top \leq \text{diag}(\boldsymbol{Z}_1^*\mathbf{Y}_1^\top)\mathbf{1}_n^\top - \boldsymbol{Z}_1^*\mathbf{Y}_1^\top .
\end{equation}
This inequality motivates Algorithm \ref{alg:cm} (see Appendix), which iteratively attempts to find a $\boldsymbol{w} \in \mathbb{R}^n$ satisfying \eqref{eq:slack_cond} up to a given tolerance. If no such $\boldsymbol{w}$ can be found after ten iterations, then the proposed value $\boldsymbol{z}_{i,1}^*$ is rejected.

Algorithm \ref{alg:cm} bears some resemblance to the Jacobi method for solving systems of linear equations, but we have yet to develop a full theoretical account of its convergence, which would allow for a more principled choice of the number of iterations needed to attain a given error tolerance. In numerical experiments, we have observed that ten iterations is sufficient to have the output of Algorithm \ref{alg:cm} agree with the output of standard methods for solving the optimal assignment problem (as implemented in the \texttt{transport} function from the \textsf{R} package \texttt{transport}).

Finally, while the discussion in this section has referred only to the first latent variable set $\boldsymbol{Z}_1$, the same steps, with different subscripts, apply to simulating $\boldsymbol{Z}_2$.

\subsection{A note on missing data}

The MCMC procedure for estimation of the CCA parameters can be modified to accommodate data missing at random (MAR). For instance, if $\boldsymbol{y}_{i, 1}$ is missing, then the multirank likelihood does not impose a cyclically monotone constraint on $\boldsymbol{z}_{i,1}$. The MCMC algorithm for simulating values of the latent $\boldsymbol{Z}_1$'s can therefore be initialized by solving an assignment problem between $\mathbf{Y}_{-i, 1}$ and $\mathbf{Z}_{-i, 1}^{(0)}$. Then, at each iteration, a new instance $\boldsymbol{z}^*_{i, 1}$ can be simulated from the unconstrained full conditional distribution
\begin{equation*}
    \boldsymbol{z}^*_{i, 1} | - \sim N_{p_1}(\boldsymbol{Q}_1 \boldsymbol{\Lambda} \boldsymbol{Q}_2^\top  \boldsymbol{z}_{i, 2}, \boldsymbol{I}_{p_1} - \boldsymbol{Q}_1 \boldsymbol{\Lambda}^2 \boldsymbol{Q}_1^\top).
\end{equation*}
If multiple rows in either $\boldsymbol{Y}_1$ or $\boldsymbol{Y}_2$ are missing, these may be treated analogously.

\section{Examples}

\subsection{Estimation of multivariate dependence in simulated data}

To illustrate how our model for semiparametric CCA extends existing models, we compare estimates derived from our method for semiparametric CCA to those of standard CCA and Gaussian copula-based CCA, where the latter corresponds to the family of models proposed by \cite{zoh_pcan_2016, agniel_analysis_2017, yoon_sparse_2020}. Specifically, we evaluate estimates derived from each model on data simulated from the semiparametric CCA model (\ref{eq:cmcca_mod}) using three different choices of cyclically monotone transformations.

For the first simulation scenario, we choose the positive definite, symmetric matrices 
\begin{equation}\label{eq:simcovs}
\boldsymbol{\Sigma}_1^{1/2} = \boldsymbol{\Sigma}_2^{1/2} = \left[\begin{array}{cc}
1 & 0.25 \\
0.25 & 1
\end{array}\right]
\end{equation} 
as our transformation functions, so the resulting data are jointly multivariate normal. In the second simulation, we apply the monotone function 
\begin{equation*}
g(z) = (\mathcal{W}^{-1}_{1,1} \circ \Phi) (z)
\end{equation*}
coordinate-wise to each variable set, where $\mathcal{W}^{-1}_{1,1}$ denotes the quantile function of the Weibull distribution with shape and scale parameter equal to $1$, and $\Phi$ denotes the standard normal CDF. The resulting data have a distribution that is a Gaussian copula, but is no longer multivariate normal. Finally, in the third simulation we apply the cyclically monotone function
\begin{equation*}
G(\boldsymbol{z}) =\boldsymbol{U}^\top (\mathcal{W}^{-1}_{1,1} \circ \Phi) (\boldsymbol{U}\boldsymbol{z})
\end{equation*}
to each variable set, where the entries of $\boldsymbol{U} \in \mathbb{R}^{2 \times 2}$ are simulated independently from $N(0, 0.75^2)$ and $\mathcal{W}^{-1}_{1,1} \circ \Phi$ is applied coordinate-wise. Note that because $\boldsymbol{U}$ is not diagonal, $G$ is not coordinate-wise monotone. Therefore, the corresponding data have a distribution that is neither normal nor a Gaussian copula, but is in our model for semiparametric CCA. 

For each simulation scenario, we evaluate four estimation methods: classical CCA, Gaussian copula-based CCA, and semiparametric CCA using the pseudolikelihood strategy of Section 3.1 and using the MCMC algorithm of Section 3.2. We compare each estimate of the CCA parameters to the true CCA parameters using the loss criterion
\begin{equation*}
L_{\boldsymbol{W}}(\hat{\boldsymbol{W}}) = \|\hat{\boldsymbol{W}} - \boldsymbol{W}\|_2^2
\end{equation*}
where $\boldsymbol{W} = \boldsymbol{Q}_1 \boldsymbol{\Lambda} \boldsymbol{Q}_2^\top$. The exact methods for computing each estimate are as follows:

\paragraph{Standard CCA}
\begin{enumerate}[noitemsep]
    \item Compute 
    \begin{enumerate}[noitemsep]
        \item[] $\hat{\boldsymbol{\Sigma}}_1 = \frac{1}{n}(\boldsymbol{Y}_1 - \frac{1}{n}\boldsymbol{1} \boldsymbol{1}^\top \boldsymbol{Y}_1)^\top 
    (\boldsymbol{Y}_1 - \frac{1}{n}\boldsymbol{1} \boldsymbol{1}^\top \boldsymbol{Y}_1)$
    \item[] $\hat{\boldsymbol{\Sigma}}_2 = \frac{1}{n}(\boldsymbol{Y}_2 - \frac{1}{n}\boldsymbol{1} \boldsymbol{1}^\top \boldsymbol{Y}_2)^\top 
    (\boldsymbol{Y}_2 - \frac{1}{n}\boldsymbol{1} \boldsymbol{1}^\top \boldsymbol{Y}_2)$
    \item[] $\hat{\boldsymbol{\Omega}} = \frac{1}{n}(\boldsymbol{Y}_1 - \frac{1}{n}\boldsymbol{1} \boldsymbol{1}^\top \boldsymbol{Y}_1)^\top 
    (\boldsymbol{Y}_2 - \frac{1}{n}\boldsymbol{1} \boldsymbol{1}^\top \boldsymbol{Y}_2)$
    \end{enumerate}
    \item Set $\hat{\boldsymbol{W}} = \hat{\boldsymbol{\Sigma}}_1^{-1/2} \hat{\boldsymbol{\Omega}} \hat{\boldsymbol{\Sigma}}_2^{-1/2}$
\end{enumerate}

\paragraph{Gaussian copula-based CCA}
\begin{enumerate}[noitemsep]
    \item Simulate 500 posterior samples of $\boldsymbol{C}$, the $(p_1 + p_2) \times (p_1 + p_2)$-dimensional correlation matrix parametrizing a Gaussian copula using the \textsf{R} package \texttt{sbgcop}.
    \item For each posterior sample, set $\boldsymbol{\Sigma}^{(t)}_1, \boldsymbol{\Sigma}^{(t)}_2, \boldsymbol{\Omega}^{(t)}$ to be the upper $p_1 \times p_1$ block, lower $p_2 \times p_2$ block, and $p_1 \times p_2$ off-diagonal block, respectively, of $\boldsymbol{C}^{(t)}$.
    \item Compute $\boldsymbol{W}^{(t)} = {\boldsymbol{\Sigma}^{(t)}_1}^{-1/2} \boldsymbol{\Omega}^{(t)} {\boldsymbol{\Sigma}^{(t)}_2}^{-1/2}$ for each posterior sample.
    \item Compute posterior mean $\hat{\boldsymbol{W}} = \frac{1}{T} \sum_{t=1}^T \boldsymbol{W}^{(t)}$.
\end{enumerate}

\paragraph{Semiparametric CCA using pseudolikelihood}\label{par:pseudo}
\begin{enumerate}[noitemsep]
    \item Simulate matrices $\boldsymbol{Z}_1, \boldsymbol{Z}_2$ with i.i.d. standard normal entries.
    \item Permute the rows of $\boldsymbol{Z}_1$ and $\boldsymbol{Z}_2$ so that the permuted matrices $\hat{\boldsymbol{Z}}_1$ and $\hat{\boldsymbol{Z}}_2$ are in cycically monotone correspondence with $\boldsymbol{Y}_1, \boldsymbol{Y}_2$, respectively.
    \item Compute 
    \begin{enumerate}[noitemsep]
        \item[] $\hat{\boldsymbol{\Sigma}}_1 = \frac{1}{n}(\hat{\boldsymbol{Z}}_1 - \frac{1}{n}\boldsymbol{1} \boldsymbol{1}^\top \hat{\boldsymbol{Z}}_1)^\top 
    (\hat{\boldsymbol{Z}}_1 - \frac{1}{n}\boldsymbol{1} \boldsymbol{1}^\top \hat{\boldsymbol{Z}}_1)$
    \item[] $\hat{\boldsymbol{\Sigma}}_2 = \frac{1}{n}(\hat{\boldsymbol{Z}}_2 - \frac{1}{n}\boldsymbol{1} \boldsymbol{1}^\top \hat{\boldsymbol{Z}}_2)^\top 
    (\hat{\boldsymbol{Z}}_2 - \frac{1}{n}\boldsymbol{1} \boldsymbol{1}^\top \hat{\boldsymbol{Z}}_2)$
    \item[] $\hat{\boldsymbol{\Omega}} = \frac{1}{n}(\hat{\boldsymbol{Z}}_1 - \frac{1}{n}\boldsymbol{1} \boldsymbol{1}^\top \hat{\boldsymbol{Z}}_1)^\top 
    (\hat{\boldsymbol{Z}}_2 - \frac{1}{n}\boldsymbol{1} \boldsymbol{1}^\top \hat{\boldsymbol{Z}}_2)$
    \end{enumerate}
    \item Set $\hat{\boldsymbol{W}} = \hat{\boldsymbol{\Sigma}}_1^{-1/2} \hat{\boldsymbol{\Omega}} \hat{\boldsymbol{\Sigma}}_2^{-1/2}$
\end{enumerate}

\paragraph{Semiparametric CCA using multirank likelihood and MCMC}
\begin{enumerate}[noitemsep]
    \item Simulate 500 posterior samples of the CCA parameters $\boldsymbol{Q}_1, \boldsymbol{Q}_2, \boldsymbol{\Lambda}$.
    \item Compute $\boldsymbol{W}^{(t)} = \boldsymbol{Q}^{(t)} _1 \boldsymbol{\Lambda}^{(t)}  \boldsymbol{Q}^{(t)\top} _2$ for each posterior sample.
    \item Compute posterior mean $\hat{\boldsymbol{W}} = \frac{1}{T} \sum_{t=1}^T \boldsymbol{W}^{(t)}$.
\end{enumerate}

\begin{figure}
\begin{subfigure}{0.5\textwidth}
\centering
\includegraphics[scale=0.48]{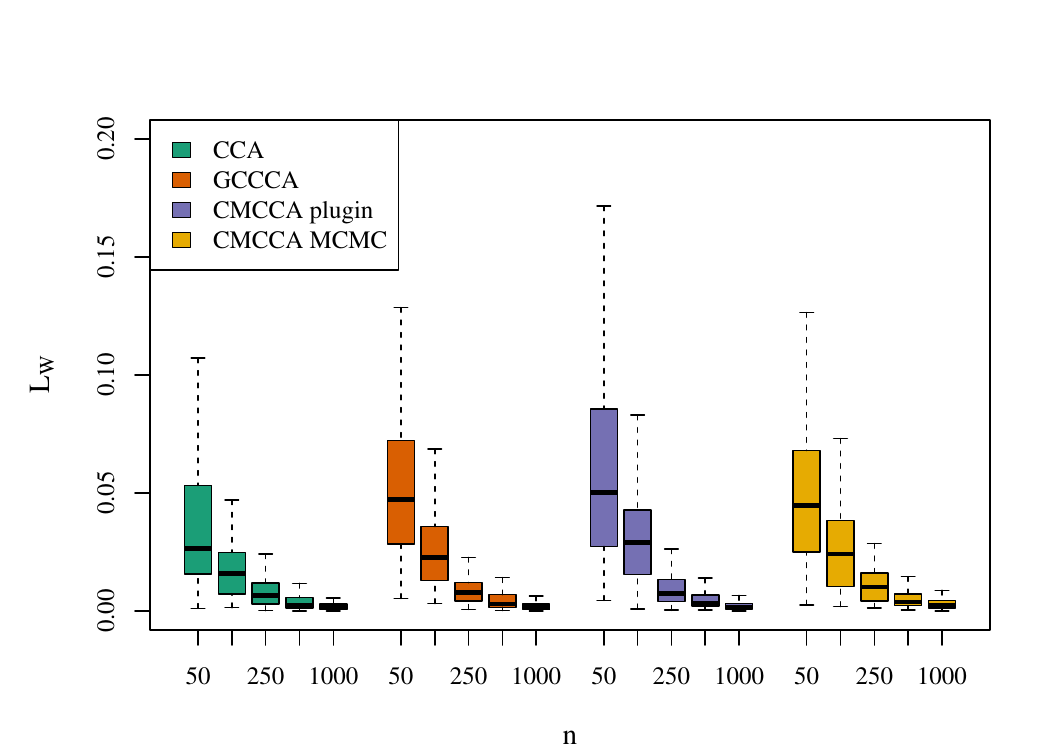}
\caption{}
\label{fig:sub1}
\end{subfigure}
\begin{subfigure}{0.5\textwidth}
\centering
\includegraphics[scale=0.48]{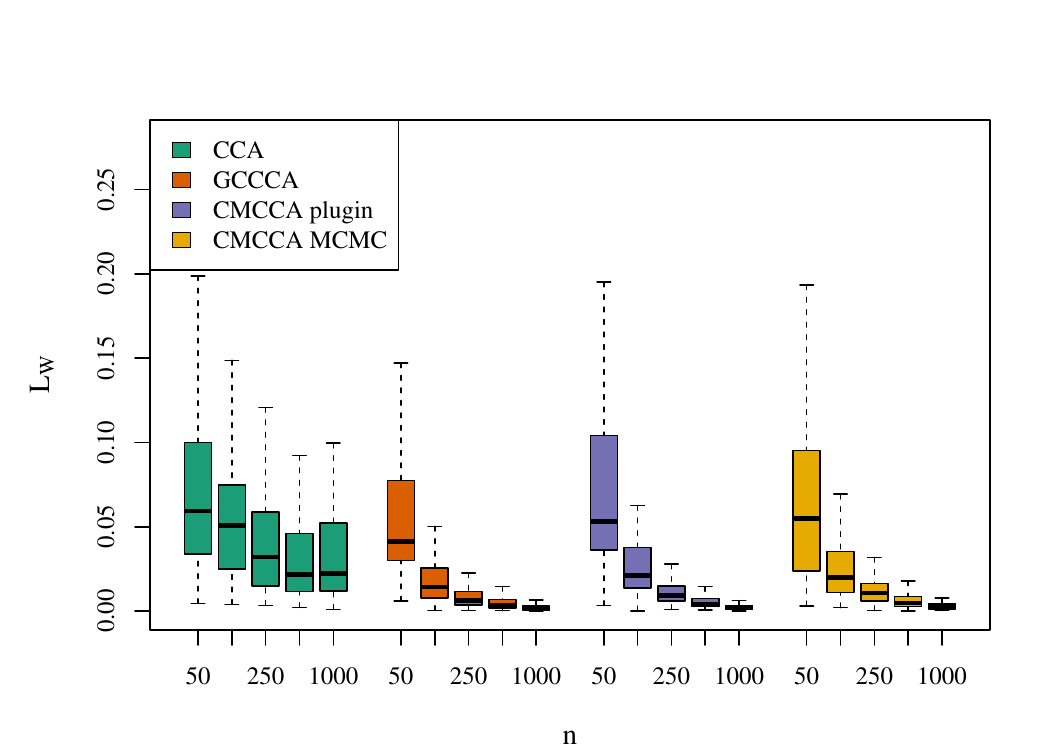}
\caption{}
\label{fig:sub2}
\end{subfigure}\newline
\begin{subfigure}{0.5\textwidth}
\centering
\includegraphics[scale=0.48]{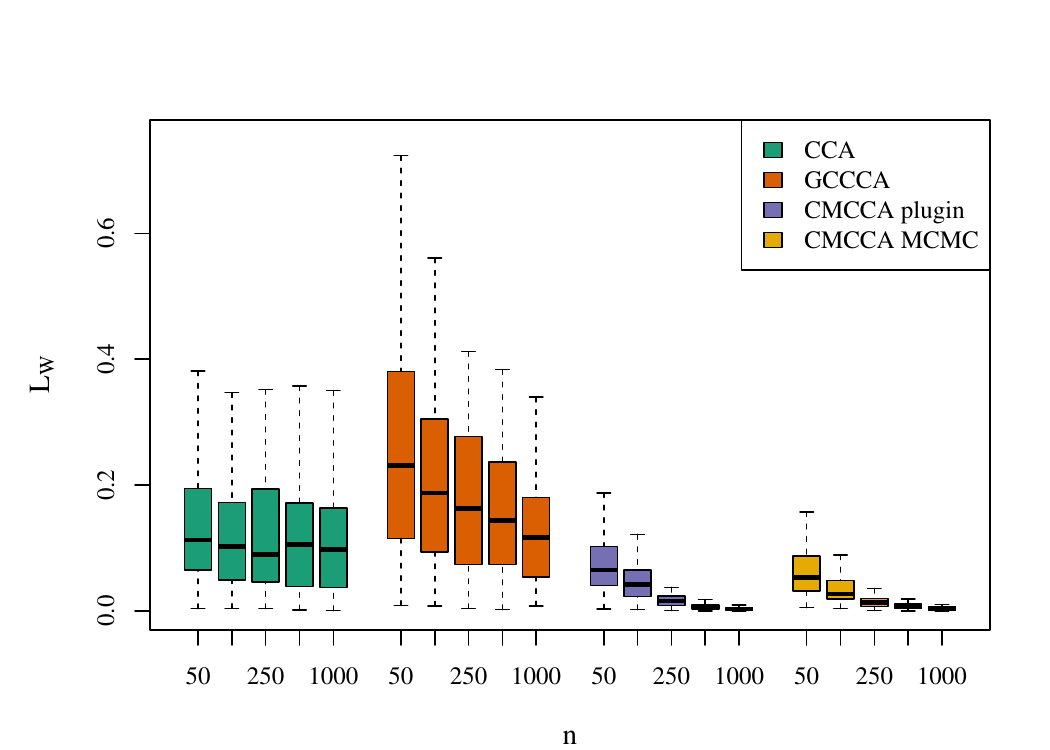}
\caption{}
\label{fig:sub3}
\end{subfigure}
\begin{subfigure}{0.5\textwidth}
\centering
\includegraphics[scale=0.48]{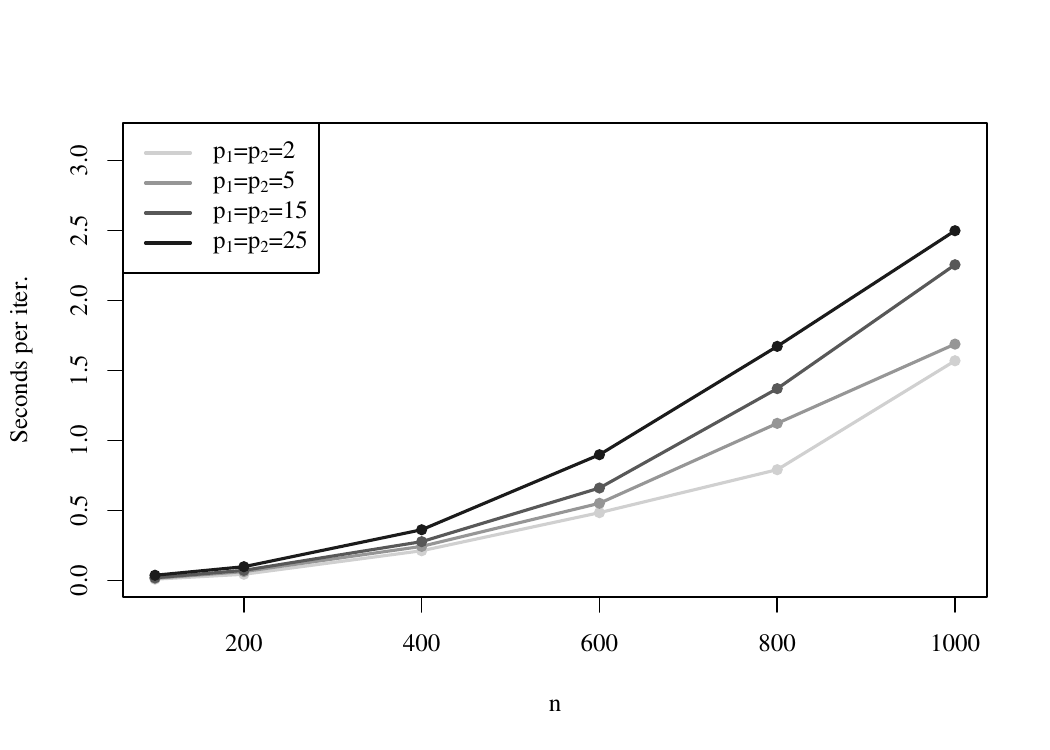}
\caption{}
\label{fig:sub4}
\end{subfigure}
\caption{Sum of squares error for three simulation scenarios and four estimation methods: traditional CCA (CCA), Gaussian copula-based CCA (GCCCA), and our methods for semiparametric CCA using the pseudolikelihood strategy of Section 3.1 (CMCCA plugin) and the algorithm of Section 3.2 (CMCCA MCMC). (a) Estimation improves with sample size for all methods. (b) Estimation with traditional CCA stops improving as $n$ increases. (c) The estimates derived from our model for semiparametric CCA improve as $n$ increases while the others either do not improve or lag behind. (d) Per-iteration run time of our CMCCA MCMC algorithm for several sample sizes and variable set dimensions.}
\label{fig:test}
\end{figure}

In Figure \ref{fig:test}, we show the distribution of $L_{\boldsymbol{W}}(\hat{\boldsymbol{W}})$ for each estimate over $100$ trials for different sample sizes. Within each trial, we first simulate $\boldsymbol{Q}_1, \boldsymbol{Q}_2, \boldsymbol{\Lambda}$ from the uniform prior distributions described in the previous section, and then we simulate $\boldsymbol{Y}_1, \boldsymbol{Y}_2$ according to \eqref{eq:cmcca_mod} using these parameter values. Hence, the means of the distributions shown represent the average risk of each estimate with respect to the uniform prior distributions on the CCA parameters. In the first scenario (Figure \ref{fig:sub1}), the estimation of all three methods improves with sample size. This is expected since the data-generating distribution, the multivariate normal, falls in each of the three model families. In the second scenario, estimation with standard CCA fails to improve with sample size because the data-generating distribution is not normal. However, the data-generating distribution for the second scenario is in the Gaussian copula model and our model for semiparametric CCA, so estimation with each of the corresponding methods improves with sample size. Finally, in the third scenario estimation with standard CCA fails to improve and estimation with Gaussian copula-based CCA lags behind as $n$ gets large. Our model for semiparametric CCA produces estimates that improve with sample size in all three simulation scenarios and clearly outperforms the other estimates in the third scenario.

Results for analogous simulations with $p_1=p_2=3$, and $p_1=p_2=5$ are presented in Figures \ref{fig:testp3} and \ref{fig:testp5} in the Appendix. For the first simulation scenario, we apply $3$- and $5$-dimensional linear transformations of the form \eqref{eq:simcovs}, with diagonal entries equal to $1$ and all off-diagonal entries equal to $0.25$. For the third simulation scenario, the $\boldsymbol{U}$ matrices have entries simulated independently from $N(0, 1/3)$ and $N(0, 1/5)$, respectively. In the $p_1=p_2=5$ simulations, we see evidence that the our semiparametric estimates suffer from an efficiency gap relative to traditional CCA and GCCCA for the first and second simulation scenarios, respectively. This increased sensitivity to dimension was also observed in \cite{shi_distribution-free_2020} and \cite{zhang_projective_2023}, where the latter proposes solutions to this issue, though not through the use of transport-based ranks.

Figure \ref{fig:sub4} displays the per-iteration run time of our implementation of the MCMC algorithm for several sample sizes and variable set dimensions. While increasing the variable set dimension scales the simulation time of the $\boldsymbol{Q}_1, \boldsymbol{Q}_2$ roughly as $\min(p_1, p_2)^3$ due to singular value decompositions, the total computation time is dominated by the simulation of the latent variables $\boldsymbol{Z}_1, \boldsymbol{Z}_2$, which requires checking cyclical monotonicity at each step. Simulations like the ones in this section that require many runs of the MCMC algorithm take several hours for $n = 1000$. However, single runs of the MCMC on smaller datasets like those analyzed in the next section can be done in several minutes.

\subsection{Analysis of multivariate dependence in two datasets}

Here we illustrate the use of our model for semiparametric CCA by analyzing multivariate dependence in two datasets. In addition to reporting the magnitude of dependence between variable sets in each dataset, we describe the nature of that dependence through quantitative and qualitative summaries of the posterior distribution of the CCA parameters using the multirank likelihood.

In the first analysis, we consider the association between year-to-year fluctuations of two sets of climate variables measured in five geographic regions of Brazil between 1961 and 2019. These data come from the Brazilian National Institute of Meteorology (INMET) and are available at \url{kaggle.com/datasets}. The first group of climate variables consists of temperature, atmospheric pressure, evaporation, insolation, and wind velocity. The second group consists of relative humidity, cloudiness, and precipitation. Individual measurements of the climate variables are averaged across weather stations and across days to obtain average monthly values for each variable within each geographic region. We then take the log ratio between the average value for a given month and the average value in the same month of the previous year to remove seasonal dependence. These log ratios are the observation unit for the climate variables in the analysis that follows.

To determine whether our model for semiparametric CCA should be preferred over traditional CCA or Gaussian copula-based CCA for these data, we apply a \textit{normal scores transformation} to each climate variable. The normal scores transformation works by aligning the empirical quantiles of each variable with that of the standard normal distribution, transforming data with arbitrary univariate margins to have approximately standard normal margins. If the data have an empirical distribution that is approximately multivariate normal after a normal scores transformation, it would be reasonable to conclude that their distribution can be described by a Gaussian copula. As reported in Table \ref{tab:hz}, even after the normal scores transformation the climate variables in this sample are not approximately jointly normally distributed according to the Henze-Zirkler test for multivariate normality \citep{henze_class_1990}. However, after applying a multivariate analog of the normal scores transformation, the climate variables appear to be approximately multivariate normal, indicating that the data may be described by our model for semiparametric CCA. Letting $\boldsymbol{Y}_1, \boldsymbol{Y}_2$ denote the matrices containing the two climate variable sets, we calculate these multivariate normal scores using the following steps:

\paragraph{Multivariate normal scores}
\begin{enumerate}[noitemsep]
    \item Simulate matrices $\boldsymbol{Z}_1, \boldsymbol{Z}_2$ with i.i.d. standard normal entries.
    \item Obtain the optimal assignments between $\boldsymbol{Z}_1, \boldsymbol{Y}_1$ and $\boldsymbol{Z}_2, \boldsymbol{Y}_2$.
    \item Permute the rows of $\boldsymbol{Z}_1$ and $\boldsymbol{Z}_2$ to obtain the multivariate normal scores $\hat{\boldsymbol{Z}}_1$ and $\hat{\boldsymbol{Z}}_2$, which are in cycically monotone correspondence with $\boldsymbol{Y}_1, \boldsymbol{Y}_2$, respectively.
\end{enumerate}

\begin{table}[h] 
\centering
\begin{tabular}{lrrrrr} 
  \hline
 & Region 1 & Region 2 & Region 3 & Region 4 & Region 5 \\ 
  \hline
H-Z statistic (normal scores) & $1.396$ & $1.318$ & $1.362$ & $1.393$ & $0.948$ \\ 
  p-value (normal scores) & $<10^{-7}$ & $<10^{-7}$ & $<10^{-7}$ & $<10^{-7}$ & $0.464$ \\ 
  H-Z statistic (m.v. normal scores) & $0.940$ & $0.985$ & $0.995$ & $0.977$ & $0.931$ \\ 
  p-value (m.v. normal scores) & $0.915$ & $0.200$ & $0.106$ & $0.273$ & $0.707$ \\ 
   \hline
\end{tabular}
\caption{Results of Henze-Zirkler test for multivariate normality on regional climate fluctuations after applying a normal scores transformation (top two rows) and after applying a multivariate normal scores transformation (bottom two rows).}
\label{tab:hz}
\end{table}

We apply our method for semiparametric CCA to each region's climate data by simulating $10,000$ values from a Markov Chain with stationary distribution equal to the posterior distribution of the CCA parameters using the multirank likelihood. We discard the first $1,000$ values and retain every $5^{\text{th}}$ subsequent iterate. Histograms depicting the approximate posterior distributions of $\lambda_1, \lambda_2$, and $\lambda_3$ for each geographic region as well as tables reporting posterior means and $95\%$ credible intervals for $\lambda_1, \lambda_2$, and $\lambda_3$ can be found in the Appendix. The upper and lower endpoints of the credible intervals are computed, respectively, as the $2.5\%$ and $97.5\%$ sample quantiles of the MCMC iterates. For all five regions, the lower endpoint of the credible interval for $\lambda_1$ is greater than $0.5$, which offers strong evidence that the two climate variable sets are not independent in any of the five regions. However, as the lower endpoints of the credible intervals for $\lambda_3$ in regions 4, and 5 are nearly equal to zero, the dependence there might reasonably be described by a two dimensional model. By contrast, the dependence in regions 1, 2, and 3 likely requires 3 dimensions to describe, as the posterior distributions of $\lambda_3$ for these are concentrated away from zero.

\begin{figure}[h]
\begin{subfigure}{.5\linewidth}
\centering
\includegraphics[scale=0.5]{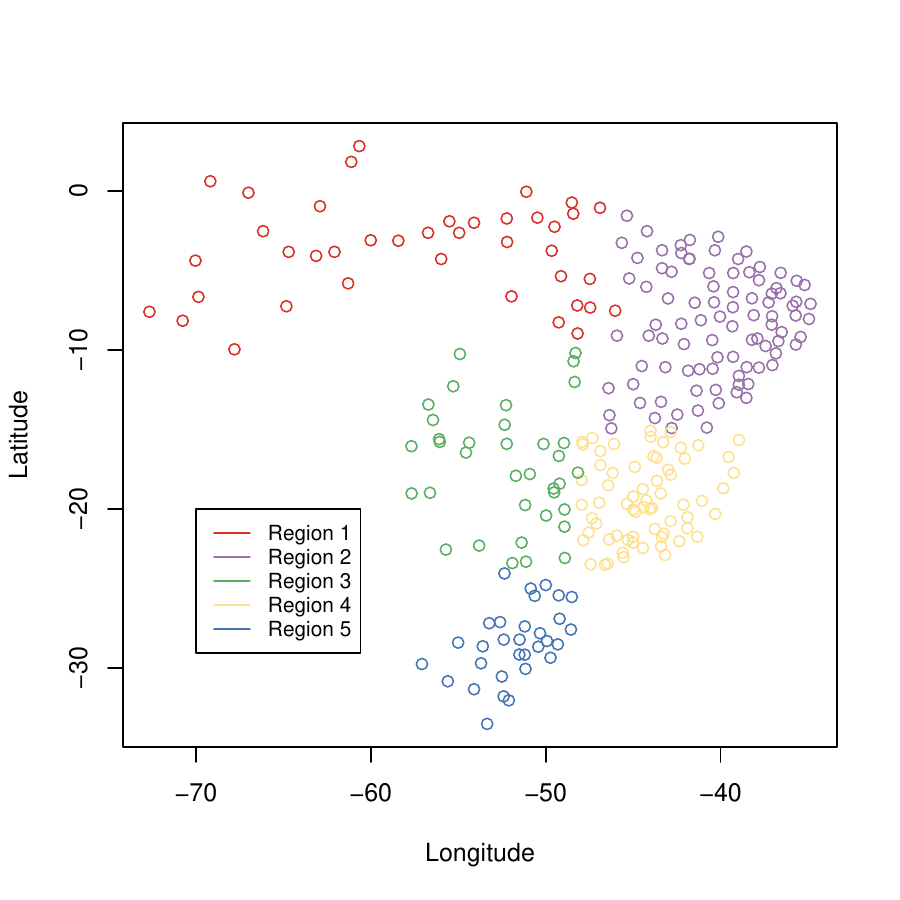}
\caption{}
\label{fig:brazilreg}
\end{subfigure}%
\begin{subfigure}{.5\linewidth}
\centering
\includegraphics[scale=0.5]{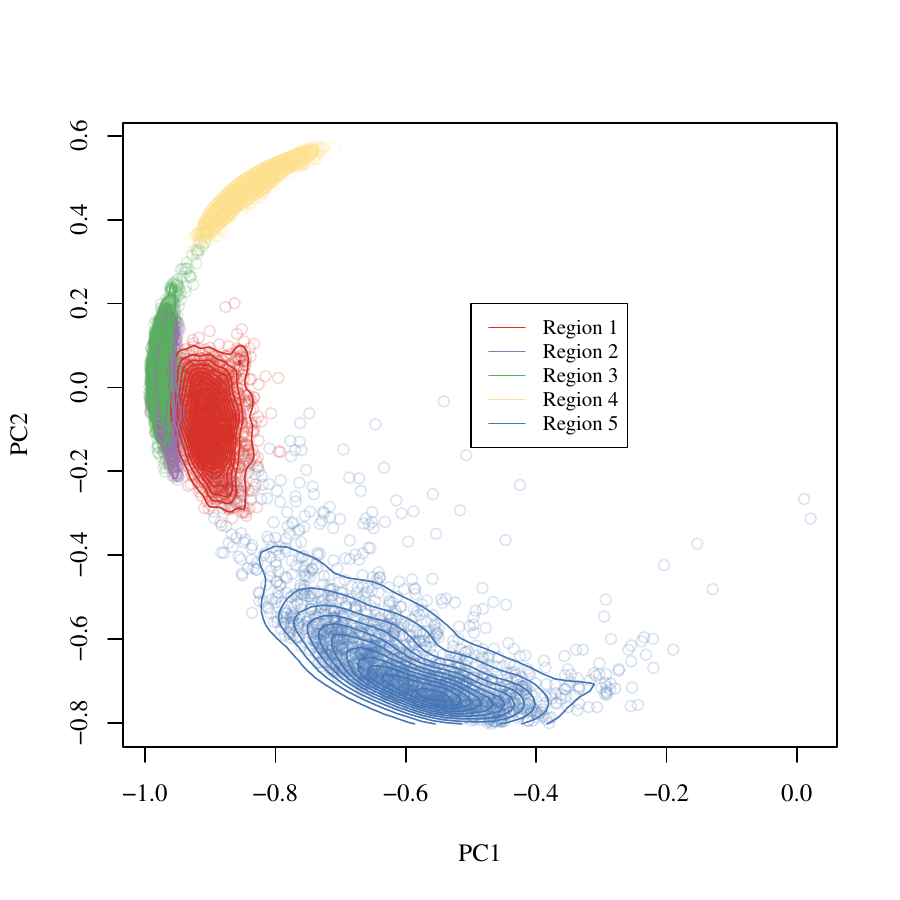}
\caption{}
\label{fig:brazilpca}
\end{subfigure}\\[1ex]
\caption{(a) Geographic locations of weather stations where climate variables were measured, colored by region. (b) Visualization of the posterior distribution of dependence parameters associated with the first canonical correlation for each region.}
\label{fig:test2}
\end{figure}

To further characterize the nature of the dependence between the sets of climate fluctuations, we analyze posterior realizations of $\boldsymbol{O}_1 = \boldsymbol{q}_{1,1} \boldsymbol{q}_{2, 1}^\top$, the outer product between the first columns of $\boldsymbol{Q}_1$ and $\boldsymbol{Q}_2$, for each geographic region. As can be seen in the decomposition
\begin{equation*}
    \boldsymbol{Q}_1 \boldsymbol{\Lambda} \boldsymbol{Q}_2^\top = \sum_{k=1}^d \lambda_k \boldsymbol{q}_{1,k} \boldsymbol{q}_{2, k}^\top = \sum_{k=1}^d \lambda_k \boldsymbol{O}_k,
\end{equation*}
$\boldsymbol{O}_1$ represents the cross-correlation structure associated with the first canonical correlation. In Figure \ref{fig:test2}, we display the geographic locations of the weather stations where the climate data were recorded (left) alongside a 2-dimensional principle components projection of the 15-dimensional posterior samples of $\text{vec}(\boldsymbol{O}_1)$ (right). The posterior distributions indicate that the associations between the climate variable sets in Regions 4 and 5 are quite distinct although they are geographically close, whereas the associations between climate variable sets in Regions 2 and 3 are very similar. A quantitative description of the similarities between the regional dependence structures is found in Table \ref{tab:brazil_sim} in the Appendix, where we report posterior summaries of the cosine similarity between the first canonical axes of each region, defined as
\begin{equation*}
    \text{CS}(\boldsymbol{O}^{r_j}_1, \boldsymbol{O}^{r_k}_1) = \text{tr}({\boldsymbol{O}^{r_j}_1}^\top \boldsymbol{O}^{r_k}_1) = ({\boldsymbol{q}^{r_j}_{1, 1}}^\top \boldsymbol{q}^{r_k}_{1, 1})({\boldsymbol{q}^{r_j}_{2, 1}}^\top \boldsymbol{q}^{r_k}_{2, 1})
\end{equation*}
for regions $r_j$ and $r_k$. Note that this is the same as the inner product used to produce the principle components visualization in Figure \ref{fig:brazilpca}.

The second dataset we analyze, downloaded from \url{finance.yahoo.com}, contains monthly log returns of adjusted closing prices of stocks in the materials and communications market sectors from 1985 to 2019. As in \cite{shi_distribution-free_2020}, we are interested in detecting dependence between the two sectors. Additionally, we seek to characterize how that dependence differs among three major economic eras in the United States: the period of deregulation and Reaganomics from 1985-1991, the period of technological growth and globalization from 1992-2007, and the aftermath of the Great Recession from 2008-2019. The stocks chosen to represent the materials market sector as classified by Global Industry Classification Standard (GICS) are DuPont de Nemours, Inc.\ (DD), Olin Corporation (OLN), BHP Group Limited (BHP), and PPG Industries, Inc.\ (PPG). The stocks chosen to represent the communications sector are Lumen Technologies, Inc.\ (LUMN), AT\&T Inc. (T), Verizon Communications Inc.\ (VZ), and Comcast Corporation (CMCSA).

\begin{figure}[h]
\centering
\includegraphics[scale=0.55]{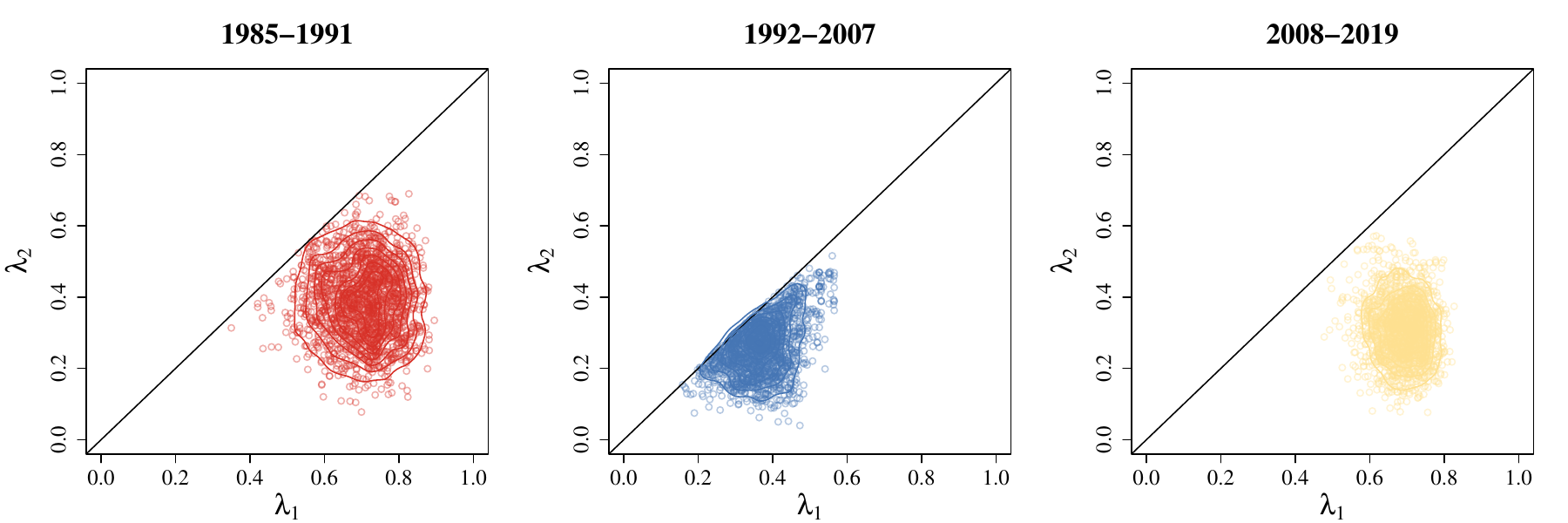}
\caption{Scatter plots of posterior realizations of $(\lambda_1, \lambda_2)$ for each of three economic eras.}
\label{fig:stock1}
\end{figure}

Posterior realizations of the CCA parameters for each economic era are obtained with MCMC as in the previous analysis. The posterior distributions of $(\lambda_1, \lambda_2)$ in Figure \ref{fig:stock1} indicate that there is dependence between the telecommunications and materials market sectors in all three economic eras. However, they also indicate that this dependence is weaker in the period from 1992-2007 than during the other eras. Moreover, the concentration of the 1992-2007 posterior distribution near the $\lambda_1 = \lambda_2$ line also indicates that the ordering of the first two canonical correlations is somewhat ambiguous; that is, the first two canonical variables explain roughly the same amount of dependence between the variable sets in this era. This contrasts to the posterior distributions of $(\lambda_1, \lambda_2)$ in the other economic eras, where clearer separation between the values of the first two canonical correlations is observed.

To understand the change in the nature of the dependence across economic eras, it may be of interest to interpret how the original variables relate to the canonical variables during each economic era. In traditional applications of CCA, the so-called \textit{canonical loadings}---the correlation coefficients between each original variable and its corresponding canonical variable---are often reported. As a marginally distribution-free analog, we report the posterior distribution of the sample Spearman rank correlation between each individual stock and its corresponding first canonical variable. Specifically, Table \ref{tab:mats} reports the posterior means and $95\%$ credible intervals of
\begin{equation}
\rho_{1,k} = \text{Cor}(R(\boldsymbol{y}_{1, k}), R(\boldsymbol{Z}_{1} \boldsymbol{q}_{1,1})),~k=1, \dots, p_1
\end{equation}
for the materials stocks, where $R(\cdot)$ yields the univariate ranks, $\text{Cor}(\cdot)$ is the Pearson correlation, and $\boldsymbol{y}_{1, k}$ denotes the $k^\text{th}$ column of $\boldsymbol{Y}_1$. Table \ref{tab:comms} reports the corresponding statistic for the communications stocks.

\begin{table}[h] 
\centering
\begin{tabular}{lllll}
  \hline
 & DD & OLN & BHP & PPG \\ 
  \hline
1985-1991 & -0.82, [-0.91, -0.68] & -0.52, [-0.75, -0.29] & -0.20, [-0.50, 0.10] & -0.55, [-0.75, -0.31] \\ 
  1992-2007 & 0.49, [-0.10, 0.88] & 0.50, [-0.08, 0.91] & 0.56, [0.01, 0.93] & 0.49, [-0.14, 0.90] \\ 
  2008-2019 & -0.78, [-0.89, -0.64] & -0.44, [-0.64, -0.25] & -0.70, [-0.83, -0.55] & -0.79, [-0.89, -0.69] \\ 
   \hline
\end{tabular}
\caption{Sample Spearman correlations (posterior means and $95\%$ credible intervals) between materials stocks and the first canonical variable of the materials stocks variable set.}
\label{tab:mats}
\end{table}

\begin{table}[h] 
\centering
\begin{tabular}{lllll}
  \hline
 & LUMN & T & VZ & CMCSA \\ 
  \hline
1985-1991 & -0.62, [-0.80, -0.40] & -0.36, [-0.62, -0.10] & -0.38, [-0.66, -0.13] & -0.83, [-0.93, -0.71] \\ 
  1992-2007 & 0.55, [-0.05, 0.92] & 0.42, [-0.11, 0.88] & 0.49, [-0.09, 0.90] & 0.36, [-0.20, 0.86] \\ 
  2008-2019 & -0.60, [-0.76, -0.44] & -0.42, [-0.64, -0.22] & -0.40, [-0.59, -0.21] & -0.79, [-0.88, -0.67] \\ 
   \hline
\end{tabular}
\caption{Sample Spearman correlations (posterior means and $95\%$ credible intervals) between communications stocks and the first canonical variable of the communications stocks variable set.}
\label{tab:comms}
\end{table}

The results in Tables \ref{tab:mats} and \ref{tab:comms} support the notion that the nature of the dependence between the materials and communications market sectors differs among the three economic eras. They also provide evidence that the weaker dependence from 1992-2007 is driven by overall market movement: the wide credible intervals indicate that the first cannonical variable of each variable set does not have a significant association to any one stock in particular. By contrast, in the years before 1992 and after 2007, individual stocks such as DD and CMCSA show stronger association to the canonical variables of the first and second variable sets, respectively.

\section{Discussion}

In this article, we have proposed a semiparametric model for CCA based on cyclically monotone transformations of jointly normal sets of variables. In contrast to existing models for CCA, our model allows the variable sets to have arbitrary multivariate marginal distributions, while retaining a parametric model of dependence. We proposed two inference strategies for the CCA parameters, the most novel of which is based on the multirank likelihood, a type of marginal likelihood, which does not depend on the multivariate marginal distributions of the variable sets. Instead, the multirank likelihood uses information in the cyclically monotone correspondence between the observed data and the unobserved normal latent variables. To simulate values subject to the multirank likelihood's cyclically monotone constraint, we have also introduced an MCMC algorithm, which allows us to use the multirank likelihood to obtain both point estimates and confidence regions for the CCA parameters.

Depending on one's philosophical inclinations, the posterior distribution over the CCA parameters using the multirank likelihood represents an exact (i.e.\ non-asymptotic) quantification of uncertainty, which may be preferable to classical standard errors that rely on normality or large sample sizes. However, interpreting the CCA parameters in our model for semiparametric CCA presents some challenges beyond those of traditional CCA. In our model, the canonical correlations $\boldsymbol{\Lambda}$ may be interpreted as magnitudes of multivariate dependence, just as in traditional CCA. However, the columns of $\boldsymbol{Q}_1, \boldsymbol{Q}_2$ correspond to the latent variable sets $\boldsymbol{Z}_1, \boldsymbol{Z}_2$, and explicitly linking these to the original variables in a manner analogous to the canonical coefficients of traditional CCA requires estimating the transformations $G_1, G_2$. While this is a viable approach, in keeping with the spirit of the rest of the article we have chosen to offer ways of interpreting the CCA parameters without respect to $G_1, G_2$. First, we have demonstrated how the columns of $\boldsymbol{Q}_1, \boldsymbol{Q}_2$ can be compared across populations to obtain similarity measures between those populations, which reflect the type of multivariate dependence observed. Second, we have proposed marginally distribution-free alternatives to canonical loadings, which, like their traditional counterparts, help to interpret the relationship between the original variables and the corresponding canonical variables.

There are several directions of research that could extend this work. For example, much of the recent interest in CCA has been driven by applications in genomics and neuroimaging, for which the variable sets are often high-dimensional compared to the sample size. In these settings, traditional CCA may fail, so it is necessary to use regularized versions of CCA. In principle our approach to semiparametric CCA is amenable to extensions allowing for Bayesian inference in such high-dimensional settings. For instance, regularized CCA could be implemented via shrinkage- or sparsity-inducing prior distributions on the entries of $\boldsymbol{\Lambda}$. Alternatively, one could enforce a low-dimensional model of dependence by combining our model with the so-called ``spiked" transport model of \cite{nilesweed_estimation_2022}. Indeed, such tools may be necessary to combat the decrease in efficiency observed for methods like ours as the dimensions of the variable sets increase (see Appendix Figure \ref{fig:testp5} or \citep{shi_distribution-free_2020, zhang_projective_2023} for discussions of related phenomena).

However, scaling the simulation methodology presented here to settings with high sample size or high dimension remains a challenge. While estimates based off of the pseudolikelihood for the CCA model can be obtained reasonably cheaply, estimates derived from our MCMC algorithm require significant computation. While subsampling schemes and certain generalized Gibbs steps \cite{liu_generalised_2000} can be incorporated into our current algorithm, our investigations so far have led us to the conclusion that achieving significantly faster computation will likely require targeting an approximation to the posterior distribution induced by the multirank likelihood, rather than the exact posterior. To this end, the application of tools from entropically-regularized optimal transport or approximate Bayesian computation (ABC) merit further study.

Another extension of this work might obtain theoretical guarantees about the identifiability of the CCA parameters in the submodel of our model for semiparametric CCA where the multivariate marginals $P_1, P_2$ are discrete. Like many recent results concerning ranks defined using cyclical monotonicity, our result on the identifiability of the CCA parameters assumes that the cyclically monotone transformations $G_1, G_2$ are absolutely continuous, and our argument following the statement of Proposition 2 assumes margins with at least some continuous part. Intuition suggests that weak identifiability conditions may also be possible in the case of entirely discrete margins. For instance, discrete margins may imply that $\boldsymbol{Q}_1, \boldsymbol{Q}_2$ are identifiable up to some rotational ambiguity, with the degree of that ambiguity depending on the support of the discrete marginal. However, obtaining a precise statement related to this issue remains a subject of ongoing investigation. Finally, although this article has focused on semiparametric CCA, inference approaches using the multirank likelihood could be useful in other semiparametric inference problems such as semiparametric regression or hierarchical models involving cyclically monotone transformations.

Code to reproduce the figures and tables in this article, as well as software for inference with the semiparametric CCA model are available at \url{https://github.com/j-g-b/cmcca}.

\subsubsection*{Acknowledgments}

The authors gratefully acknowledge Michael Jauch for helpful comments regarding MCMC methods for orthogonal matrices.

\subsubsection*{Funding}

Jonathan Niles-Weed's work was supported by the National Science Foundation under Grant DMS-2015291.

\newpage

\section{Appendix}

\subsection{An algorithm for verifying the cyclical monotonicity condition}

\begin{algorithm}
\caption{Check cyclical monotonicity}\label{alg:cm}
\begin{algorithmic}
\Require $\mathbf{Z}, \mathbf{Y} \in \mathbb{R}^{n \times p}$, $\boldsymbol{w} := \mathbf{0}$, $\varepsilon := 10^{-12}$, $\mu_\text{curr} := -1$, $\delta := 0$ 

\State $\mathbf{S} = \text{diag}(\mathbf{Z}\mathbf{Y}^\top)\mathbf{1}_n^\top - \mathbf{Z}\mathbf{Y}^\top$
\While{$\mu_\text{curr} < -\varepsilon$}

\State $\boldsymbol{w} = \text{colMax}(\boldsymbol{w}\boldsymbol{1}_n^\top-\mathbf{S})$ \Comment{Column-wise maximum}

\State $\mu_\text{next} = \min\{\mathbf{S} - (\boldsymbol{w} \mathbf{1}_n^\top - \mathbf{1}_n \boldsymbol{w}^\top)\}$ \Comment{Element-wise minimum}

\If{$|\mu_\text{next} - \mu_\text{curr}| < \epsilon$}
    \State $\delta = \delta + 1$ \Comment{Increment count if no progress}
\Else
    \State $\delta = 0$ \Comment{Otherwise reset count}
\EndIf

\If{$\delta > 10 \And \mu_\text{next} < -\epsilon$}
    \State \Return False
\EndIf

\EndWhile
\State \Return True
\end{algorithmic}
\end{algorithm}

\subsection{Additional figures and tables}

\begin{figure}
\begin{subfigure}{0.5\textwidth}
\centering
\includegraphics[scale=0.48]{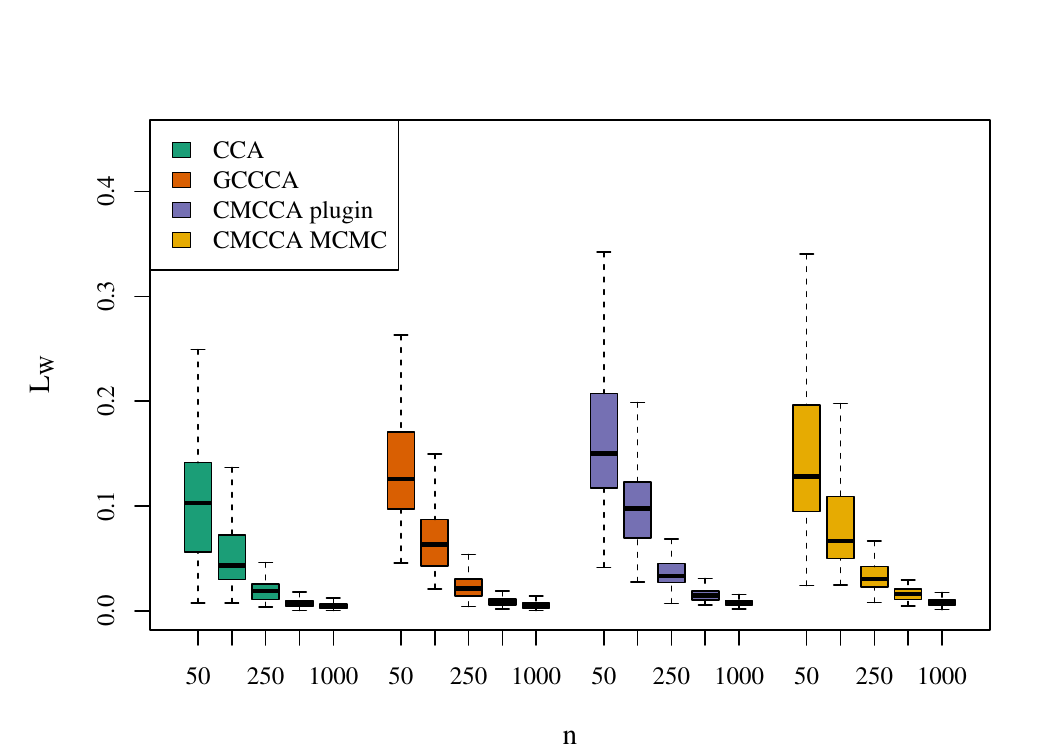}
\caption{}
\label{fig:sub1p3}
\end{subfigure}
\begin{subfigure}{0.5\textwidth}
\centering
\includegraphics[scale=0.48]{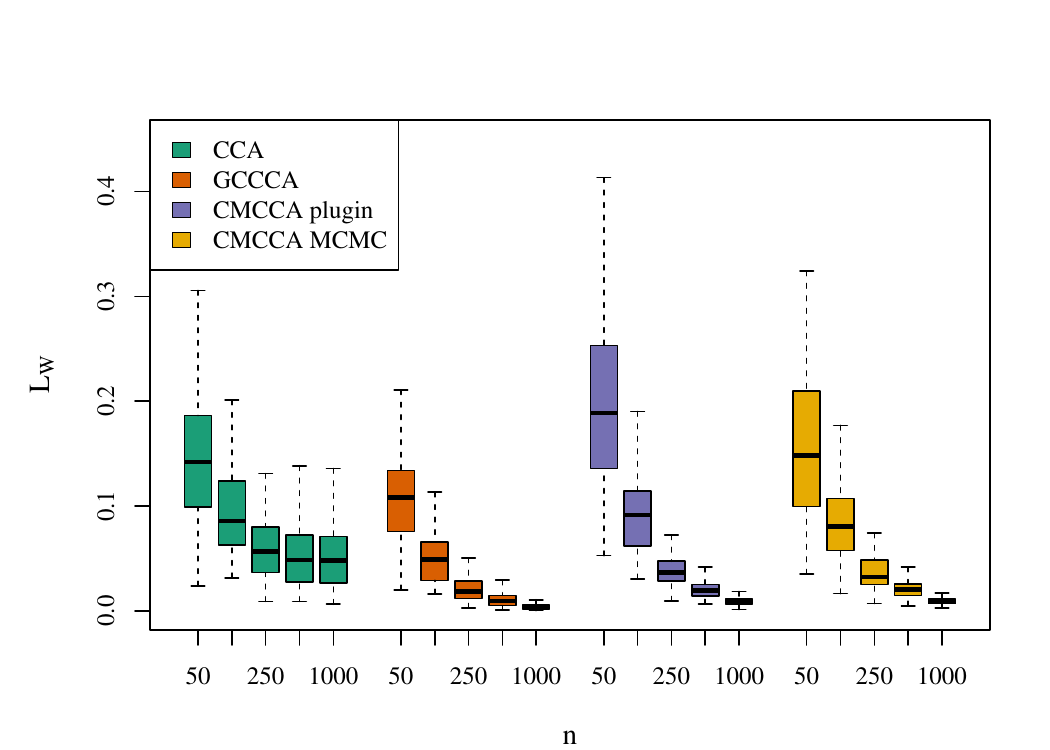}
\caption{}
\label{fig:sub2p3}
\end{subfigure}\newline
\begin{subfigure}{\textwidth}
\centering
\includegraphics[scale=0.48]{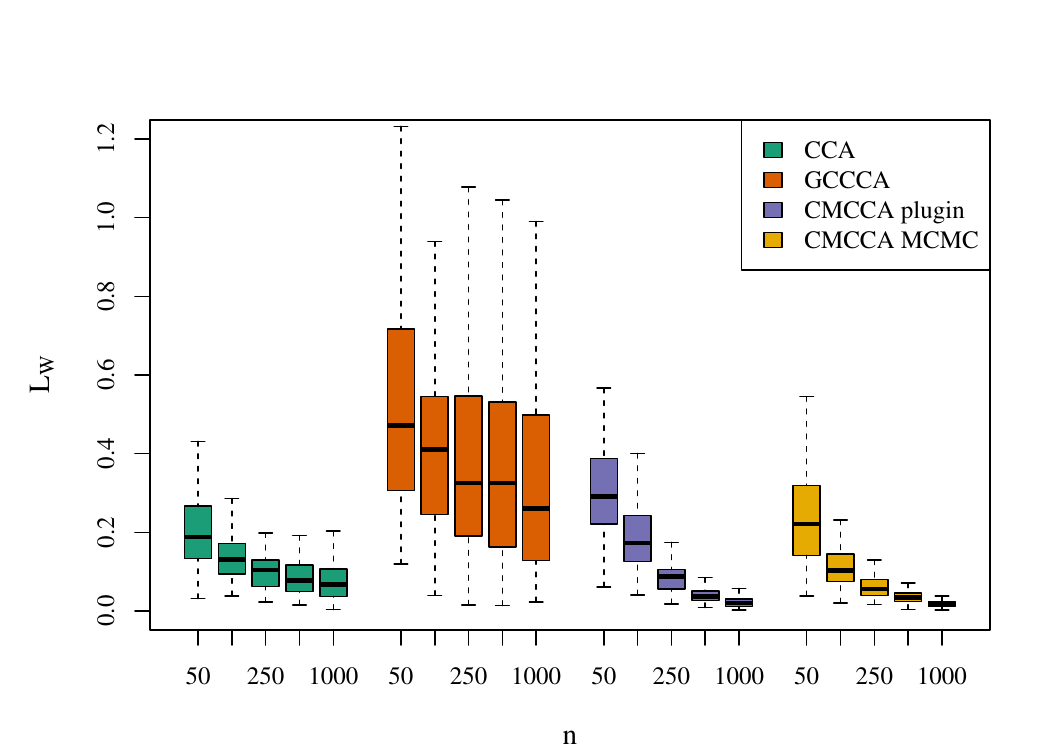}
\caption{}
\label{fig:sub3p3}
\end{subfigure}
\caption{Results of simulation study for $p_1 = p_2 = 3$. Sum of squares error for three simulation scenarios and four estimation methods: traditional CCA (CCA), Gaussian copula-based CCA (GCCCA), semiparametric CCA using the pseudolikelihood strategy of Section 3.1 (CMCCA plugin), and semiparametric CCA using the algorithm of Section 3.2 (CMCCA MCMC). (a) Estimation improves with sample size for all methods. (b) Estimation with traditional CCA stops improving as $n$ increases. (c) The estimates derived from our model for semiparametric CCA show the best improvement for $n \geq 250$.}
\label{fig:testp3}
\end{figure}

\begin{figure}
\begin{subfigure}{0.5\textwidth}
\centering
\includegraphics[scale=0.48]{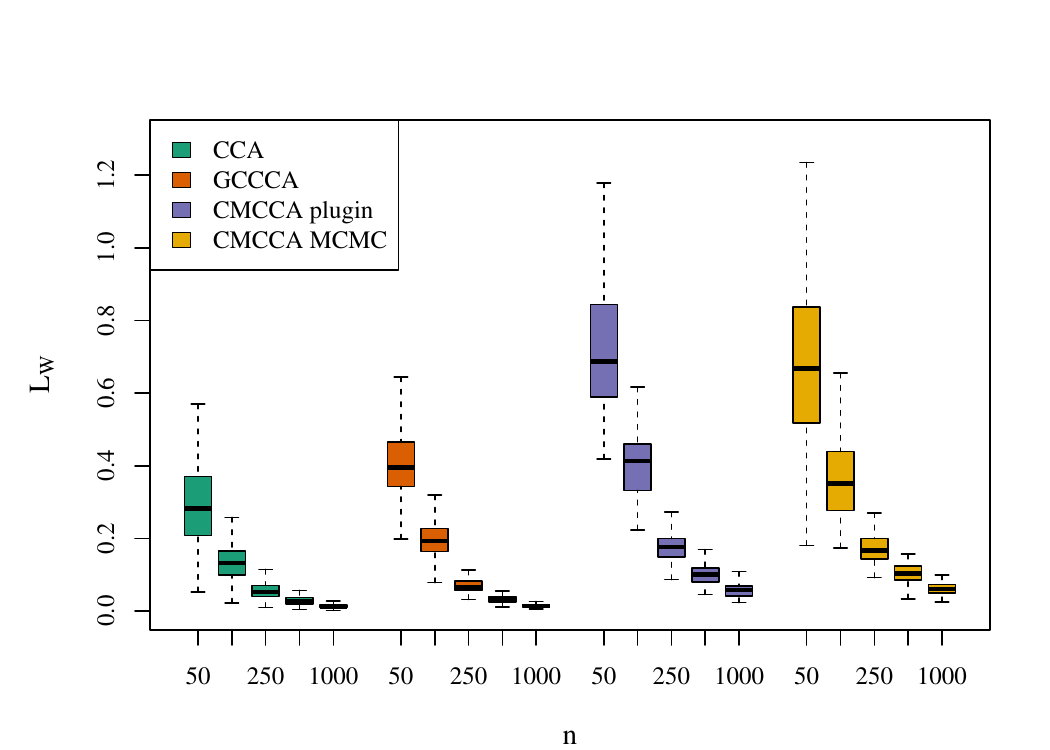}
\caption{}
\label{fig:sub1p5}
\end{subfigure}
\begin{subfigure}{0.5\textwidth}
\centering
\includegraphics[scale=0.48]{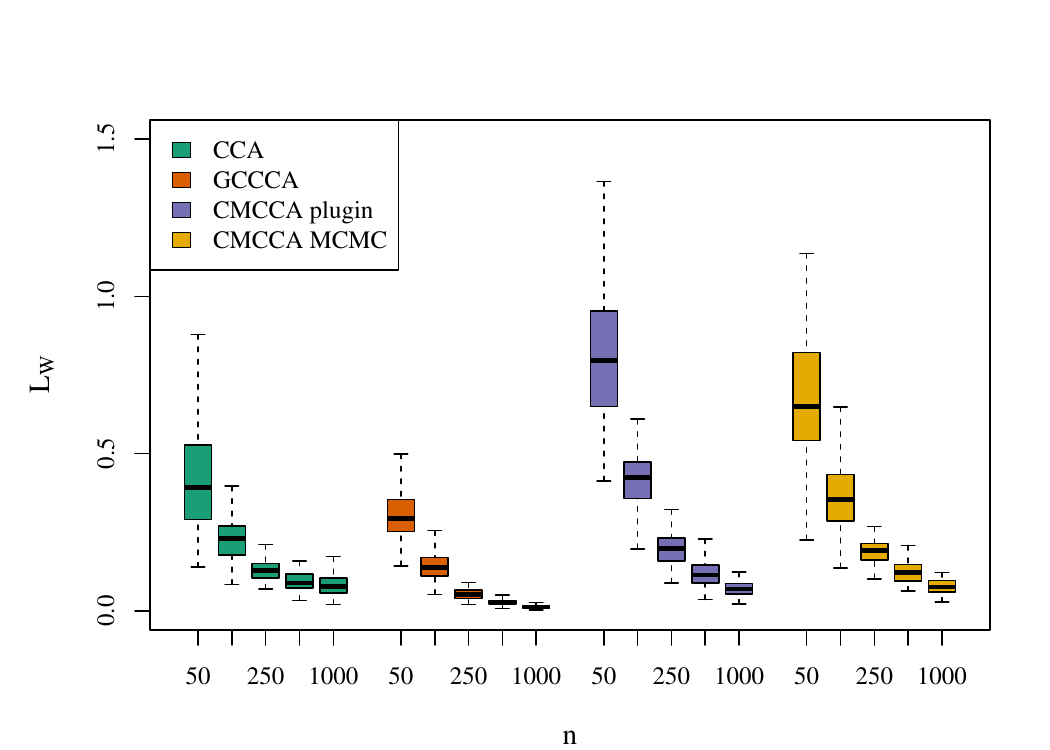}
\caption{}
\label{fig:sub2p5}
\end{subfigure}\newline
\begin{subfigure}{\textwidth}
\centering
\includegraphics[scale=0.48]{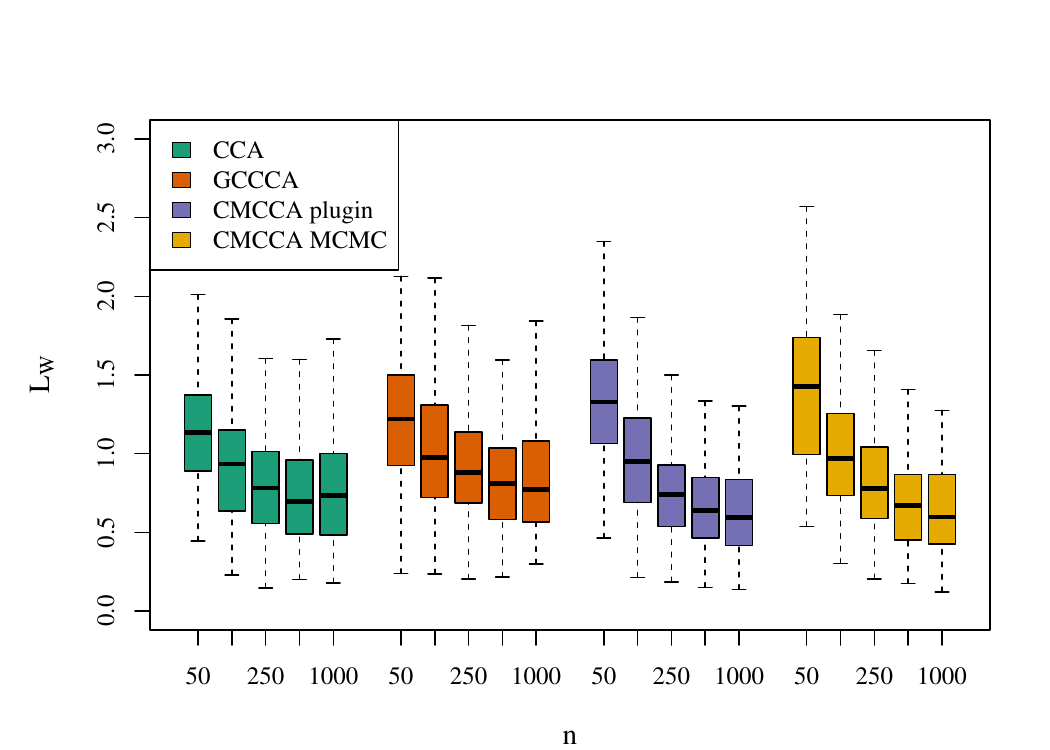}
\caption{}
\label{fig:sub3p5}
\end{subfigure}
\caption{Results of simulation study for $p_1 = p_2 = 5$. The sum of squares error for three simulation scenarios and four estimation methods: traditional CCA (CCA), Gaussian copula-based CCA (GCCCA), and our methods for semiparametric CCA using the pseudolikelihood strategy of Section 3.1 (CMCCA plugin) and the algorithm of Section 3.2 (CMCCA MCMC). (a) Estimation improves with sample size for all methods, while CCA appears most efficient. (b) Estimation improves with sample size for all methods, while GCCCA appears most efficient. (c) Only the estimates derived from our model for semiparametric CCA improve for $n \geq 500$, though the improvement is small.}
\label{fig:testp5}
\end{figure}

\begin{figure}[h]
\centering
\includegraphics[scale=0.8]{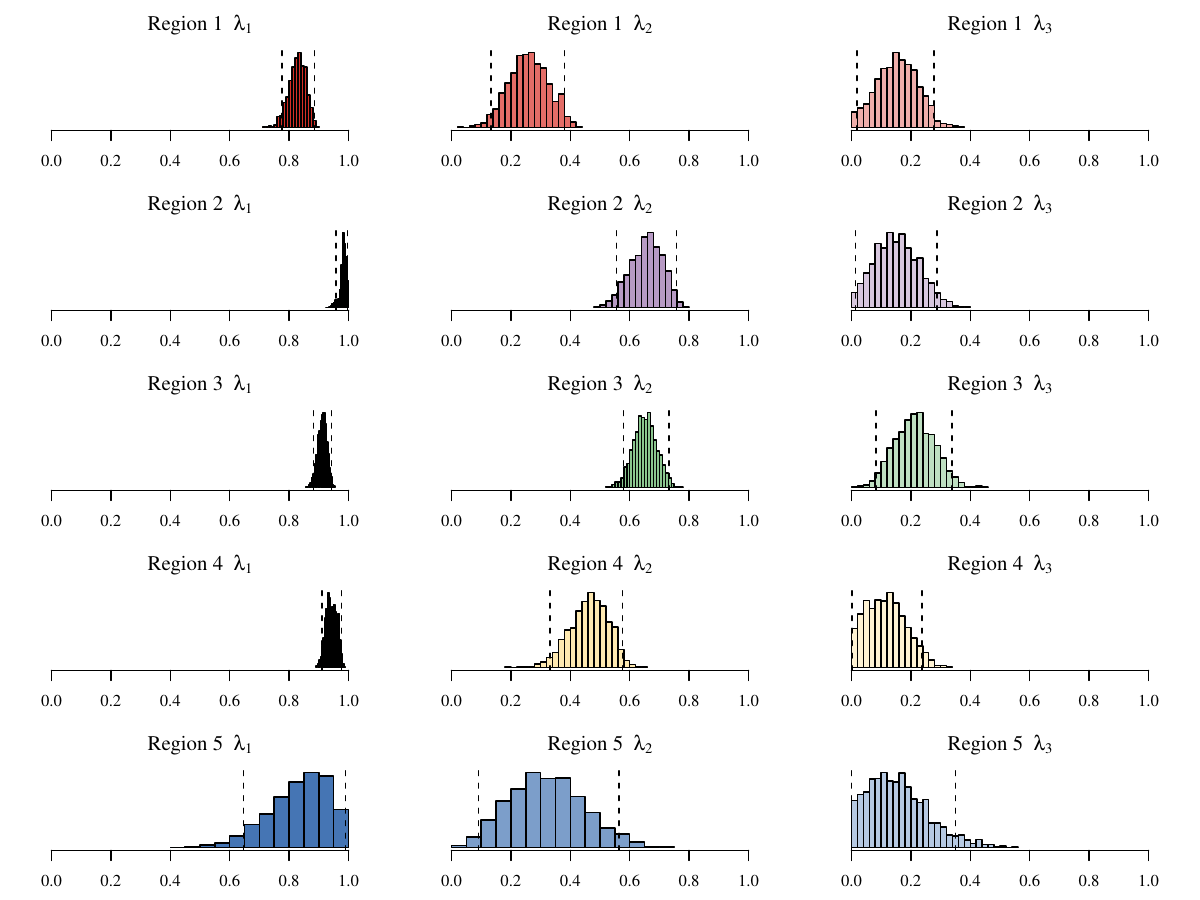}
\caption{Histograms of posterior realizations of $\lambda_1, \lambda_2, \lambda_3$ for each of the five geographic regions in Brazil. Dotted lines denote the boundaries of $95\%$ credible intervals.}
\label{fig:brazil_cancor}
\end{figure}

\begin{table}[h]
\centering
\begin{tabular}{llllll}
  \hline
 & Region 1 & Region 2 & Region 3 & Region 4 & Region 5 \\ 
  \hline
Lambda 1 & 0.83 [0.78, 0.89] & 0.98 [0.96, 0.99] & 0.91 [0.88, 0.94] & 0.94 [0.91, 0.98] & 0.83 [0.65, 0.99] \\ 
  Lambda 2 & 0.26 [0.13, 0.38] & 0.66 [0.56, 0.76] & 0.65 [0.58, 0.73] & 0.46 [0.33, 0.58] & 0.32 [0.09, 0.56] \\ 
  Lambda 3 & 0.16 [0.02, 0.28] & 0.15 [0.01, 0.29] & 0.21 [0.08, 0.34] & 0.12 [0.00, 0.24] & 0.16 [0.00, 0.35] \\ 
   \hline
\end{tabular}
\caption{Posterior means and $95\%$ credible intervals for $\lambda_1, \lambda_2, \lambda_3$ for each geographic region in Brazil.}
\end{table}

\begin{table}[h]
\centering
\begin{tabular}{lllll}
  \hline
 & Region 2 & Region 3 & Region 4 & Region 5 \\ 
  \hline
Region 1 & 0.83, [0.76, 0.91] & 0.86, [0.77, 0.94] & 0.66, [0.52, 0.81] & 0.53, [0.27, 0.77] \\ 
  Region 2 & . & 0.94, [0.89, 0.99] & 0.81, [0.686, 0.931] & 0.58, [0.31, 0.86] \\ 
  Region 3 & . & . & 0.84, [0.72, 0.95] & 0.55, [0.28, 0.83] \\ 
  Region 4 & . & . & . & 0.25, [-0.01, 0.52] \\ 
   \hline
\end{tabular}
\caption{Cosine similarity (posterior means and $95\%$ credible intervals) between the first canonical axes of climate fluctuations of each geographic region in Brazil.}
\label{tab:brazil_sim}
\end{table}

\newpage

\bibliographystyle{myjmva}
\bibliography{bib}

\begin{thebibliography}{46}
\expandafter\ifx\csname natexlab\endcsname\relax\def\natexlab#1{#1}\fi
\providecommand{\bibinfo}[2]{#2}
\ifx\xfnm\relax \def\xfnm[#1]{\unskip,\space#1}\fi
\bibitem[{Agniel and Cai(2017)}]{agniel_analysis_2017}
\bibinfo{author}{D.~Agniel}, \bibinfo{author}{T.~Cai}, \bibinfo{title}{Analysis
  of multiple diverse phenotypes via semiparametric canonical correlation
  analysis}, \bibinfo{journal}{Biom} \bibinfo{volume}{73}
  (\bibinfo{year}{2017}) \bibinfo{pages}{1254--1265}.
\bibitem[{Ambrosio and Gigli(2013)}]{ambrosio_users_2013}
\bibinfo{author}{L.~Ambrosio}, \bibinfo{author}{N.~Gigli}, \bibinfo{title}{A
  {User}’s {Guide} to {Optimal} {Transport}}, in:
  \bibinfo{booktitle}{Modelling and {Optimisation} of {Flows} on {Networks}},
  volume \bibinfo{volume}{2062}, \bibinfo{publisher}{Springer Berlin
  Heidelberg}, \bibinfo{address}{Berlin, Heidelberg}, \bibinfo{year}{2013}, pp.
  \bibinfo{pages}{1--155}.
\bibitem[{Anderson(1999)}]{anderson_asymptotic_1999}
\bibinfo{author}{T.~Anderson}, \bibinfo{title}{Asymptotic {Theory} for
  {Canonical} {Correlation} {Analysis}}, \bibinfo{journal}{Journal of
  Multivariate Analysis} \bibinfo{volume}{70} (\bibinfo{year}{1999})
  \bibinfo{pages}{1--29}.
\bibitem[{Bach and Jordan(2005)}]{bach_probabilistic_2005}
\bibinfo{author}{F.~R. Bach}, \bibinfo{author}{M.~I. Jordan}, \bibinfo{title}{A
  probabilistic interpretation of canonical correlation analysis},
  \bibinfo{type}{Technical Report}, \bibinfo{year}{2005}.
\bibitem[{Bakirov et~al.(2006)Bakirov, Rizzo and
  Székely}]{bakirov_multivariate_2006}
\bibinfo{author}{N.~K. Bakirov}, \bibinfo{author}{M.~L. Rizzo},
  \bibinfo{author}{G.~J. Székely}, \bibinfo{title}{A multivariate
  nonparametric test of independence}, \bibinfo{journal}{Journal of
  Multivariate Analysis} \bibinfo{volume}{97} (\bibinfo{year}{2006})
  \bibinfo{pages}{1742--1756}.
\bibitem[{del Barrio et~al.(2020)del Barrio, González-Sanz and
  Hallin}]{del_barrio_note_2020}
\bibinfo{author}{E.~del Barrio}, \bibinfo{author}{A.~González-Sanz},
  \bibinfo{author}{M.~Hallin}, \bibinfo{title}{A note on the regularity of
  optimal-transport-based center-outward distribution and quantile functions},
  \bibinfo{journal}{Journal of Multivariate Analysis} \bibinfo{volume}{180}
  (\bibinfo{year}{2020}) \bibinfo{pages}{104671}.
\bibitem[{Brenier(1991)}]{brenier_polar_1991}
\bibinfo{author}{Y.~Brenier}, \bibinfo{title}{Polar factorization and monotone
  rearrangement of vector-valued functions}, \bibinfo{journal}{Comm. Pure Appl.
  Math.} \bibinfo{volume}{44} (\bibinfo{year}{1991}) \bibinfo{pages}{375--417}.
\bibitem[{Chen et~al.(2013)Chen, Bushman, Lewis, Wu and
  Li}]{chen_structure-constrained_2013}
\bibinfo{author}{J.~Chen}, \bibinfo{author}{F.~D. Bushman},
  \bibinfo{author}{J.~D. Lewis}, \bibinfo{author}{G.~D. Wu},
  \bibinfo{author}{H.~Li}, \bibinfo{title}{Structure-constrained sparse
  canonical correlation analysis with an application to microbiome data
  analysis}, \bibinfo{journal}{Biostatistics} \bibinfo{volume}{14}
  (\bibinfo{year}{2013}) \bibinfo{pages}{244--258}.
\bibitem[{Chernozhukov et~al.(2017)Chernozhukov, Galichon, Hallin and
  Henry}]{chernozhukov_mongekantorovich_2017}
\bibinfo{author}{V.~Chernozhukov}, \bibinfo{author}{A.~Galichon},
  \bibinfo{author}{M.~Hallin}, \bibinfo{author}{M.~Henry},
  \bibinfo{title}{Monge–{Kantorovich} depth, quantiles, ranks and signs},
  \bibinfo{journal}{Ann. Statist.} \bibinfo{volume}{45} (\bibinfo{year}{2017}).
\bibitem[{Deb et~al.(2021)Deb, Bhattacharya and Sen}]{deb_efficiency_2021}
\bibinfo{author}{N.~Deb}, \bibinfo{author}{B.~B. Bhattacharya},
  \bibinfo{author}{B.~Sen}, \bibinfo{title}{Efficiency {Lower} {Bounds} for
  {Distribution}-{Free} {Hotelling}-{Type} {Two}-{Sample} {Tests} {Based} on
  {Optimal} {Transport}}, \bibinfo{journal}{arXiv:2104.01986 [math, stat]}
  (\bibinfo{year}{2021}). \bibinfo{note}{ArXiv: 2104.01986}.
\bibitem[{Deb and Sen(2021)}]{deb_multivariate_2021}
\bibinfo{author}{N.~Deb}, \bibinfo{author}{B.~Sen},
  \bibinfo{title}{Multivariate {Rank}-{Based} {Distribution}-{Free}
  {Nonparametric} {Testing} {Using} {Measure} {Transportation}},
  \bibinfo{journal}{Journal of the American Statistical Association}
  (\bibinfo{year}{2021}) \bibinfo{pages}{1--16}.
\bibitem[{Dempster et~al.(1977)Dempster, Laird and
  Rubin}]{dempster_maximum_1977}
\bibinfo{author}{A.~P. Dempster}, \bibinfo{author}{N.~M. Laird},
  \bibinfo{author}{D.~B. Rubin}, \bibinfo{title}{Maximum {Likelihood} from
  {Incomplete} {Data} {Via} the \textit{{EM}} {Algorithm}},
  \bibinfo{journal}{Journal of the Royal Statistical Society: Series B
  (Methodological)} \bibinfo{volume}{39} (\bibinfo{year}{1977})
  \bibinfo{pages}{1--22}.
\bibitem[{Eaton(1983)}]{eaton_multivariate_1983}
\bibinfo{author}{M.~L. Eaton}, \bibinfo{title}{Multivariate statistics: a
  vector space approach}, Wiley series in probability and mathematical
  statistics, \bibinfo{publisher}{Wiley}, \bibinfo{address}{New York},
  \bibinfo{year}{1983}.
\bibitem[{Fan and Henry(2023)}]{fan_vector_2023}
\bibinfo{author}{Y.~Fan}, \bibinfo{author}{M.~Henry}, \bibinfo{title}{Vector
  copulas}, \bibinfo{journal}{Journal of Econometrics} \bibinfo{volume}{234}
  (\bibinfo{year}{2023}) \bibinfo{pages}{128--150}.
\bibitem[{Figalli(2018)}]{figalli_continuity_2018}
\bibinfo{author}{A.~Figalli}, \bibinfo{title}{On the continuity of
  center-outward distribution and quantile functions},
  \bibinfo{journal}{Nonlinear Analysis} \bibinfo{volume}{177}
  (\bibinfo{year}{2018}) \bibinfo{pages}{413--421}.
\bibitem[{Gelfand and Smith(1990)}]{gelfand_sampling-based_1990}
\bibinfo{author}{A.~E. Gelfand}, \bibinfo{author}{A.~F.~M. Smith},
  \bibinfo{title}{Sampling-{Based} {Approaches} to {Calculating} {Marginal}
  {Densities}}, \bibinfo{journal}{Journal of the American Statistical
  Association} \bibinfo{volume}{85} (\bibinfo{year}{1990})
  \bibinfo{pages}{398--409}.
\bibitem[{Genest et~al.(1995)Genest, Ghoudi and
  Rivest}]{genest_semiparametric_1995}
\bibinfo{author}{C.~Genest}, \bibinfo{author}{K.~Ghoudi},
  \bibinfo{author}{L.-P. Rivest}, \bibinfo{title}{A semiparametric estimation
  procedure of dependence parameters in multivariate families of
  distributions}, \bibinfo{journal}{Biometrika} \bibinfo{volume}{82}
  (\bibinfo{year}{1995}) \bibinfo{pages}{543--552}.
\bibitem[{Ghosal and Sen(2022)}]{ghosal_multivariate_2022}
\bibinfo{author}{P.~Ghosal}, \bibinfo{author}{B.~Sen},
  \bibinfo{title}{Multivariate ranks and quantiles using optimal transport:
  {Consistency}, rates and nonparametric testing}, \bibinfo{journal}{Ann.
  Statist.} \bibinfo{volume}{50} (\bibinfo{year}{2022}).
\bibitem[{Hallin(2017)}]{hallin_distribution_2017}
\bibinfo{author}{M.~Hallin}, \bibinfo{title}{On {Distribution} and {Quantile}
  {Functions}, {Ranks} and {Signs} in {R}\_d}, \bibinfo{type}{Working {Papers}
  {ECARES}} \bibinfo{number}{ECARES 2017-34}, ULB – Universite Libre de
  Bruxelles, \bibinfo{year}{2017}.
\bibitem[{Hallin et~al.(2021)Hallin, Del~Barrio, Cuesta-Albertos and
  Matrán}]{hallin_distribution_2021}
\bibinfo{author}{M.~Hallin}, \bibinfo{author}{E.~Del~Barrio},
  \bibinfo{author}{J.~Cuesta-Albertos}, \bibinfo{author}{C.~Matrán},
  \bibinfo{title}{Distribution and quantile functions, ranks and signs in
  dimension d: {A} measure transportation approach}, \bibinfo{journal}{Ann.
  Statist.} \bibinfo{volume}{49} (\bibinfo{year}{2021}).
\bibitem[{Hallin et~al.(2023)Hallin, La~Vecchia and
  Liu}]{hallin_rank-based_2023}
\bibinfo{author}{M.~Hallin}, \bibinfo{author}{D.~La~Vecchia},
  \bibinfo{author}{H.~Liu}, \bibinfo{title}{Rank-based testing for
  semiparametric {VAR} models: {A} measure transportation approach},
  \bibinfo{journal}{Bernoulli} \bibinfo{volume}{29} (\bibinfo{year}{2023}).
\bibitem[{Henze and Zirkler(1990)}]{henze_class_1990}
\bibinfo{author}{N.~Henze}, \bibinfo{author}{B.~Zirkler}, \bibinfo{title}{A
  class of invariant consistent tests for multivariate normality},
  \bibinfo{journal}{Communications in Statistics - Theory and Methods}
  \bibinfo{volume}{19} (\bibinfo{year}{1990}) \bibinfo{pages}{3595--3617}.
\bibitem[{Hoff(2007)}]{hoff_extending_2007}
\bibinfo{author}{P.~D. Hoff}, \bibinfo{title}{Extending the rank likelihood for
  semiparametric copula estimation}, \bibinfo{journal}{Ann. Appl. Stat.}
  \bibinfo{volume}{1} (\bibinfo{year}{2007}) \bibinfo{pages}{265--283}.
\bibitem[{Hoff(2009)}]{hoff_simulation_2009}
\bibinfo{author}{P.~D. Hoff}, \bibinfo{title}{Simulation of the {Matrix}
  {Bingham}–von {Mises}–{Fisher} {Distribution}, {With} {Applications} to
  {Multivariate} and {Relational} {Data}}, \bibinfo{journal}{Journal of
  Computational and Graphical Statistics} \bibinfo{volume}{18}
  (\bibinfo{year}{2009}) \bibinfo{pages}{438--456}.
\bibitem[{Hoff et~al.(2014)Hoff, Niu and Wellner}]{hoff_information_2014}
\bibinfo{author}{P.~D. Hoff}, \bibinfo{author}{X.~Niu}, \bibinfo{author}{J.~A.
  Wellner}, \bibinfo{title}{Information bounds for {Gaussian} copulas},
  \bibinfo{journal}{Bernoulli} \bibinfo{volume}{20} (\bibinfo{year}{2014}).
\bibitem[{Hotelling(1936)}]{hotelling_relations_1936}
\bibinfo{author}{H.~Hotelling}, \bibinfo{title}{Relations {Between} {Two}
  {Sets} of {Variates}}, \bibinfo{journal}{Biometrika} \bibinfo{volume}{28}
  (\bibinfo{year}{1936}) \bibinfo{pages}{321}.
\bibitem[{Huang and Sen(2023)}]{huang_multivariate_2023}
\bibinfo{author}{Z.~Huang}, \bibinfo{author}{B.~Sen},
  \bibinfo{title}{Multivariate {Symmetry}: {Distribution}-{Free} {Testing} via
  {Optimal} {Transport}}, \bibinfo{year}{2023}. \bibinfo{note}{ArXiv:2305.01839
  [stat]}.
\bibitem[{Jauch et~al.(2021)Jauch, Hoff and Dunson}]{jauch_monte_2021}
\bibinfo{author}{M.~Jauch}, \bibinfo{author}{P.~D. Hoff},
  \bibinfo{author}{D.~B. Dunson}, \bibinfo{title}{Monte {Carlo} {Simulation} on
  the {Stiefel} {Manifold} via {Polar} {Expansion}}, \bibinfo{journal}{Journal
  of Computational and Graphical Statistics}  (\bibinfo{year}{2021})
  \bibinfo{pages}{1--10}.
\bibitem[{Jendoubi and Strimmer(2019)}]{jendoubi_whitening_2019}
\bibinfo{author}{T.~Jendoubi}, \bibinfo{author}{K.~Strimmer}, \bibinfo{title}{A
  whitening approach to probabilistic canonical correlation analysis for omics
  data integration}, \bibinfo{journal}{BMC Bioinformatics} \bibinfo{volume}{20}
  (\bibinfo{year}{2019}) \bibinfo{pages}{15}.
\bibitem[{Liu(2000)}]{liu_generalised_2000}
\bibinfo{author}{J.~Liu}, \bibinfo{title}{Generalised {Gibbs} sampler and
  multigrid {Monte} {Carlo} for {Bayesian} computation},
  \bibinfo{journal}{Biometrika} \bibinfo{volume}{87} (\bibinfo{year}{2000})
  \bibinfo{pages}{353--369}.
\bibitem[{Mardia et~al.(1979)Mardia, Kent and Bibby}]{mardia_multivariate_1979}
\bibinfo{author}{K.~V. Mardia}, \bibinfo{author}{J.~T. Kent},
  \bibinfo{author}{J.~M. Bibby}, \bibinfo{title}{Multivariate analysis},
  Probability and mathematical statistics, \bibinfo{publisher}{Academic Press},
  \bibinfo{address}{London ; New York}, \bibinfo{year}{1979}.
\bibitem[{McCann(1995)}]{mccann_existence_1995}
\bibinfo{author}{R.~J. McCann}, \bibinfo{title}{Existence and uniqueness of
  monotone measure-preserving maps}, \bibinfo{journal}{Duke Math. J.}
  \bibinfo{volume}{80} (\bibinfo{year}{1995}) \bibinfo{pages}{309--323}.
\bibitem[{Murray et~al.(2010)Murray, Adams and MacKay}]{murray_elliptical_2010}
\bibinfo{author}{I.~Murray}, \bibinfo{author}{R.~Adams},
  \bibinfo{author}{D.~MacKay}, \bibinfo{title}{Elliptical slice sampling}, in:
  \bibinfo{editor}{Y.~W. Teh}, \bibinfo{editor}{M.~Titterington} (Eds.),
  \bibinfo{booktitle}{Proceedings of the {Thirteenth} {International}
  {Conference} on {Artificial} {Intelligence} and {Statistics}},
  volume~\bibinfo{volume}{9} of \text{\bibinfo{series}{Proceedings of {Machine}
  {Learning} {Research}}}, \bibinfo{publisher}{PMLR}, \bibinfo{address}{Chia
  Laguna Resort, Sardinia, Italy}, \bibinfo{year}{2010}, pp.
  \bibinfo{pages}{541--548}.
\bibitem[{Niles-Weed and Rigollet(2022)}]{nilesweed_estimation_2022}
\bibinfo{author}{J.~Niles-Weed}, \bibinfo{author}{P.~Rigollet},
  \bibinfo{title}{Estimation of wasserstein distances in the spiked transport
  model}, \bibinfo{journal}{Bernoulli} \bibinfo{volume}{28}
  (\bibinfo{year}{2022}) \bibinfo{pages}{2663--2688}.
\bibitem[{Oakes(1994)}]{oakes_multivariate_1994}
\bibinfo{author}{D.~Oakes}, \bibinfo{title}{Multivariate survival
  distributions}, \bibinfo{journal}{Journal of Nonparametric Statistics}
  \bibinfo{volume}{3} (\bibinfo{year}{1994}) \bibinfo{pages}{343--354}.
\bibitem[{Pan et~al.(2020)Pan, Wang, Zhang, Zhu and Zhu}]{pan_ball_2020}
\bibinfo{author}{W.~Pan}, \bibinfo{author}{X.~Wang},
  \bibinfo{author}{H.~Zhang}, \bibinfo{author}{H.~Zhu},
  \bibinfo{author}{J.~Zhu}, \bibinfo{title}{Ball {Covariance}: {A} {Generic}
  {Measure} of {Dependence} in {Banach} {Space}}, \bibinfo{journal}{Journal of
  the American Statistical Association} \bibinfo{volume}{115}
  (\bibinfo{year}{2020}) \bibinfo{pages}{307--317}.
\bibitem[{Pauli et~al.(2011)Pauli, Racugno and Ventura}]{pauli_bayesian_2011}
\bibinfo{author}{F.~Pauli}, \bibinfo{author}{W.~Racugno},
  \bibinfo{author}{L.~Ventura}, \bibinfo{title}{Bayesian composite marginal
  likelihoods}, \bibinfo{journal}{Statistica Sinica} \bibinfo{volume}{21}
  (\bibinfo{year}{2011}) \bibinfo{pages}{149--164}.
\bibitem[{Peyré and Cuturi(2018)}]{peyre_computational_2018}
\bibinfo{author}{G.~Peyré}, \bibinfo{author}{M.~Cuturi},
  \bibinfo{title}{Computational {Optimal} {Transport}},
  \bibinfo{journal}{arXiv:1803.00567 [stat]}  (\bibinfo{year}{2018}).
  \bibinfo{note}{ArXiv: 1803.00567}.
\bibitem[{Rockafellar(1966)}]{rockafellar_characterization_1966}
\bibinfo{author}{R.~Rockafellar}, \bibinfo{title}{Characterization of the
  subdifferentials of convex functions}, \bibinfo{journal}{Pacific J. Math.}
  \bibinfo{volume}{17} (\bibinfo{year}{1966}) \bibinfo{pages}{497--510}.
\bibitem[{Severini(2000)}]{severini_likelihood_2000}
\bibinfo{author}{T.~A. Severini}, \bibinfo{title}{Likelihood methods in
  statistics}, number~\bibinfo{number}{22} in \bibinfo{series}{Oxford
  statistical science series}, \bibinfo{publisher}{Oxford University Press},
  \bibinfo{address}{Oxford ; New York}, \bibinfo{year}{2000}.
\bibitem[{Sherry and Henson(2005)}]{sherry_conducting_2005}
\bibinfo{author}{A.~Sherry}, \bibinfo{author}{R.~K. Henson},
  \bibinfo{title}{Conducting and {Interpreting} {Canonical} {Correlation}
  {Analysis} in {Personality} {Research}: {A} {User}-{Friendly} {Primer}},
  \bibinfo{journal}{Journal of Personality Assessment} \bibinfo{volume}{84}
  (\bibinfo{year}{2005}) \bibinfo{pages}{37--48}.
\bibitem[{Shi et~al.(2020)Shi, Drton and Han}]{shi_distribution-free_2020}
\bibinfo{author}{H.~Shi}, \bibinfo{author}{M.~Drton}, \bibinfo{author}{F.~Han},
  \bibinfo{title}{Distribution-{Free} {Consistent} {Independence} {Tests} via
  {Center}-{Outward} {Ranks} and {Signs}}, \bibinfo{journal}{Journal of the
  American Statistical Association}  (\bibinfo{year}{2020})
  \bibinfo{pages}{1--16}.
\bibitem[{Winkler et~al.(2020)Winkler, Renaud, Smith and
  Nichols}]{winkler_permutation_2020}
\bibinfo{author}{A.~M. Winkler}, \bibinfo{author}{O.~Renaud},
  \bibinfo{author}{S.~M. Smith}, \bibinfo{author}{T.~E. Nichols},
  \bibinfo{title}{Permutation inference for canonical correlation analysis},
  \bibinfo{journal}{NeuroImage} \bibinfo{volume}{220} (\bibinfo{year}{2020})
  \bibinfo{pages}{117065}.
\bibitem[{Yoon et~al.(2020)Yoon, Carroll and Gaynanova}]{yoon_sparse_2020}
\bibinfo{author}{G.~Yoon}, \bibinfo{author}{R.~J. Carroll},
  \bibinfo{author}{I.~Gaynanova}, \bibinfo{title}{Sparse semiparametric
  canonical correlation analysis for data of mixed types},
  \bibinfo{journal}{Biometrika} \bibinfo{volume}{107} (\bibinfo{year}{2020})
  \bibinfo{pages}{609--625}.
\bibitem[{Zhang and Zhu(2023)}]{zhang_projective_2023}
\bibinfo{author}{Y.~Zhang}, \bibinfo{author}{L.~Zhu},
  \bibinfo{title}{Projective independence tests in high dimensions: the curses
  and the cures}, \bibinfo{journal}{Biometrika}  (\bibinfo{year}{2023})
  \bibinfo{pages}{asad070}.
\bibitem[{Zoh et~al.(2016)Zoh, Mallick, Ivanov, Baladandayuthapani, Manyam,
  Chapkin, Lampe and Carroll}]{zoh_pcan_2016}
\bibinfo{author}{R.~S. Zoh}, \bibinfo{author}{B.~Mallick},
  \bibinfo{author}{I.~Ivanov}, \bibinfo{author}{V.~Baladandayuthapani},
  \bibinfo{author}{G.~Manyam}, \bibinfo{author}{R.~S. Chapkin},
  \bibinfo{author}{J.~W. Lampe}, \bibinfo{author}{R.~J. Carroll},
  \bibinfo{title}{{PCAN}: {Probabilistic} correlation analysis of two
  non‐normal data sets}, \bibinfo{journal}{Biom} \bibinfo{volume}{72}
  (\bibinfo{year}{2016}) \bibinfo{pages}{1358--1368}.

\end{thebibliography}
\end{document}